\documentclass[useAMS,usenatbib]{mn2e}
\pdfoutput=1
\usepackage{aas_macros}
\usepackage{epsfig}
\usepackage{amsmath}
\usepackage{amssymb}
\usepackage{comment}
\usepackage{caption}
\usepackage{leftidx}
\usepackage{natbib}
\usepackage[rgb]{xcolor}
\definecolor{MyGreen}{rgb}{0.0,0.6,0.3}
\definecolor{MyPurple}{rgb}{0.6,0,0.3}
\usepackage{fancyref}
\usepackage{hyperref}
\hypersetup{colorlinks=true,citecolor=MyGreen,linkcolor=MyPurple,urlcolor=blue}

\def\beq{\begin{equation}}
\def\eeq{\end{equation}}
\def\ba{\begin{eqnarray}}
\def\ea{\end{eqnarray}}
\def\bal{\begin{align}}
\def\eal{\end{align}}
\def\bxi{{\mbox{\boldmath $\xi$}}}

\begin{document}

\title[Compact Object Spins in Binaries] {The spins of compact objects born from helium stars in binary systems}

\author[Fuller \& Lu]{
Jim Fuller$^{1}$\thanks{Email: jfuller@caltech.edu}, Wenbin Lu$^{2}$
\\$^1$TAPIR, Mailcode 350-17, California Institute of Technology, Pasadena, CA 91125, USA
\\$^2$Department of Astrophysical Sciences, Princeton University, Princeton, NJ 08544, USA
}

\label{firstpage}
\maketitle

\begin{abstract}

The angular momentum (AM) content of massive stellar cores helps to determine the natal spin rates of neutron stars and black holes. Asteroseismic measurements of low-mass stars have proven that stellar cores rotate slower than predicted by most prior work, so revised models are necessary. In this work, we apply an updated AM transport model based on the Tayler instability to massive helium stars in close binaries, in which tidal spin-up can greatly increase the star's AM. Consistent with prior work, these stars can produce highly spinning black holes upon core-collapse if the orbital period is less than $P_{\rm orb} \lesssim \! 1 \, {\rm day}$. For neutron stars, we predict a strong correlation between the pre-explosion mass and the neutron star rotation rate, with millisecond periods ($P_{\rm NS} \lesssim 5 \, {\rm ms}$) only achievable for massive ($M \gtrsim 10 \, M_\odot$) helium stars in tight ($P_{\rm orb} \lesssim 1 \, {\rm day}$) binaries. Finally, we discuss our models in relation to type Ib/c supernovae, superluminous supernove, gamma-ray bursts, and LIGO measurements of black hole spins. Our models are roughly consistent with the rates and energetics of these phenomena, with the exception of broad-lined Ic supernovae, whose high rates and ejecta energies are difficult to explain.


\end{abstract}


\section{Introduction}

Rotation is a key player in the drama that unfolds upon the death of a massive star. The angular momentum (AM) contained in the iron core and overlying layers determines the rotation rate at core-collapse (CC), which could have a strong impact on the dynamics of CC and the subsequent supernova (SN) \citep[see e.g.,][]{macfadyen:99,woosley:02,yoon:06}. Rotation may help determine the nature of the compact remnant, which could range from a slowly rotating neutron star (NS) to a millisecond magnetar or a rapidly spinning black hole \citep[BH, see e.g.,][]{heger:00,heger:05}. The former could evolve into an ordinary pulsar, while the latter two outcomes offer exciting prospects for the engines that power long gamma-ray bursts (GRBs), broad-lined type Ic SNe (Ic-BL), and type Ic superluminous SNe \citep{woosley:93,maeda:07,kasen:10,metzger:11}.

In most cases, however, compact objects are probably born slowly rotating due to efficient AM transport within their progenitor stars. This is demonstrated by several lines of evidence. Asteroseismic measurements of core rotation rates of red giant stars \citep{beck:12,mosser:12,deheuvels:14,deheuvels:15,triana:17,gehan:18} indicate that they rotate much slower than predicted by most AM transport models \citep{cantiello:14,fullerwave:14,belkacem:15,spada:16,eggenberger:17,ouazzani:18}. White dwarfs typically rotate with periods of $\sim 1 \, {\rm day}$ \citep{hermes:17}, proving that the vast majority of their progenitors' core AM is extracted by the time they form. Pulsar studies indicate that typical NSs are born with natal rotation periods of $P_{\rm 0} \sim 50-100 \, {\rm ms}$ \citep{faucher:06,popov:10,popov:12,gullon:14}, again slower than predicted by many models \citep{heger:05}. Finally, most BH spins measured by LIGO are low \citep{ligoo2b:18,zaldarriaga:18,miller:20,roulet:21,zevin:21}, again indicating that strong AM transport within their progenitors extracts AM from the core and transports it to the envelope where it is lost by winds, binary interactions, or a SN explosion. This appears to conflict with the high BH spins measured for high-mass X-ray binaries \citep{miller:15}, an issue which is not yet understood (see \citealt{qin:19}, \citealt{fishbach:21} and \citealt{belczynski:21} for recent discussion). 

\cite{fuller:19} argued that magnetic fields generated by the Tayler instability create magnetic torques that can extract most of a stellar core's AM and can approximately match asteroseismic observation rates (but see also \citealt{eggenberger:19}). That model is a modified version of the Tayler-Spruit dynamo \citep{spruit:02} used in many stellar models, in which weak magnetic fields are amplified by differential rotation. In the \cite{fuller:19} version, the fields grow to larger strengths (creating more efficient AM transport) due to weaker non-linear dissipation via Alfv\'en waves in the saturated state. \cite{ma:19} applied this model to massive single stars, predicting long NS rotation periods of tens to hundreds of milliseconds, while \cite{fullerma:19} predicted low BH spins ($a \sim 10^{-2}$) for BHs born from single stars.

While compact object spins are slow in most cases, central engine models for energetic supernovae and GRBs suggest that rapid core rotation is possible in a small fraction of SN progenitors. A natural possibility to explore is stripped-envelope stars born in close binaries, where tidal spin-up replenishes the AM of the helium star, potentially spawning rapidly rotating compact objects (see, e.g., \citealt{kushnir:17,qin:18,bavera:19,belczynski:20,olejak:21}). Here we investigate that possibility in detail by generating a suite of models in compact binary systems, including tidal torques and an updated AM transport prescription based on \cite{fuller:19}. We will find that rapid NS and BH spins are indeed possible, but that they are restricted to close binaries (orbital periods $P_{\rm orb} \lesssim 1 \, {\rm d}$) and are most likely to originate from very massive helium stars ($M_{\rm He} \gtrsim 5-10 \, M_\odot$).

\section{Angular Momentum Transport for Low Shear}
\label{angmom}

The Tayler-Spruit dynamo \citep{spruit:02} produces magnetic torques by generating radial magnetic field from azimuthal fields via the Tayler instability. The azimuthal field $B_\phi$ is generated by winding up the radial field $B_r$ via differential rotation. The azimuthal field grows until it becomes unstable to the Tayler instability \citep{spruit:99}, such that magnetic energy is then channeled into small-scale structures. The Tayler instability saturates when non-linear damping processes dissipate energy at the same rate that energy is added to the azimuthal field by winding. When the instability operates, it may also produce a magnetic dynamo that alters the mean radial magnetic field strength.

In prior descriptions of the Tayler instability \citep{spruit:02,fuller:19}, it is not clear what happens when the shear $q = |d \ln \Omega/d \ln r|$ is below the minimum value $q_{\rm min}$ needed to sustain the saturated state, where $r$ is radius and $\Omega$ is the local rotation rate of the star. According to the model of \cite{spruit:02}, $q_{\rm min} = (N_{\rm eff}/\Omega)^{7/4} (\eta/r^2 \Omega)^{1/4}$, where $N_{\rm eff}$ is the effective buoyancy frequency and $\eta$ is the magnetic diffusivity, while $q_{\rm min} \sim (N_{\rm eff}/\Omega)^{5/2} (\eta/r^2 \Omega)^{3/4}$ in the model of \cite{fuller:19}, and both models assume that $\Omega \ll N_{\rm eff}$. When $q=q_{\rm min}$, the azimuthal magnetic field has an Alfv\'en frequency $\omega_{\rm A} = B_\phi/\sqrt{4 \pi \rho r^2}$ that is equal to the minimum field strength needed for Tayler instability, $\omega_c \sim \Omega (N_{\rm eff}/\Omega)^{1/2} (\eta/r^2 \Omega)^{1/4}$.

When $q < q_{\rm min}$, energy is added to the background field at a rate too slow to keep up with non-linear damping when the instability operates. The radial magnetic field $B_r$ will still be wound up into a toroidal field $B_\phi$ by differential rotation, increasing the Alfv\'en frequency until it reaches $\omega_{\rm A} \sim \omega_c$. However, for $q < q_{\rm min}$, the shear is not strong enough to maintain the saturated states found by \cite{spruit:02} or \cite{fuller:19}, so the instability likely occurs intermittently. 

To estimate the torque provided by the intermittent action of the Tayler instability, we posit that the time-averaged torque is
\beq
T \sim \frac{q}{q_{\rm min}} B_r B_\phi
\eeq
when $q < q_{\rm min}$ and $T \sim B_r B_\phi$ otherwise.
This corresponds to a torque of the same magnitude as the saturated state, but acting on a duty cycle $q/q_{\rm min}$. Assuming that $B_r \sim (\omega_{\rm A}/N) B_\phi$ as used by both \cite{fuller:19} and \cite{spruit:02}, the time-averaged torque can be written
\beq
T \sim 4 \pi \rho r^2 \frac{q}{q_{\rm min}} \frac{\omega_{\rm A}^3}{N} \, .
\eeq


When the instability occurs intermittently, the time-averaged value of $\omega_{\rm A}$ will be very close to the threshold $\omega_c$, since field strengths above this will initiate the instability which will decrease $B_\phi$ faster than it can be generated due to winding. Using $q_{\rm min} \sim \alpha^{-3} (N/\Omega)^{5/2} (\eta/r^2 \Omega)^{3/4}$ from \cite{fuller:19}, where $\alpha$ is a constant of order unity, and the value of $\omega_c$ above, we find that the torque evaluates to
\begin{equation}
T \sim 4 \pi \alpha^3 \rho r^2 q \Omega^4/N^2 \, ,
\end{equation}
i.e., the same value found when the instability operates continuously when $q > q_{\rm min}$. Hence, the associated viscosity can always be written
\begin{equation}
\nu_{\rm AM} = \alpha^3 r^2 \frac{\Omega^3}{N^2} \, ,
\end{equation}
the same as found by \cite{fuller:19}. The reason that the torque can reach the same value even when the instability operates intermittently is that the time-averaged value of $B_\phi$ is larger than it would be in the saturated state from \cite{fuller:19}. In other words, when $q < q_{\rm min}$, $\omega_{\rm A} \sim \omega_c > \Omega (q \Omega/N)^{1/3}$, where the last value is the saturated value of $\omega_{\rm A}$ from \cite{fuller:19}.

Although we recover the same viscosity law $\nu = \alpha^3 r^2 \Omega^3/N^2$, this now applies to the entire radiative region of the star where $\Omega < N$, not just where $q > q_{\rm min}$. Hence, the value of $\alpha$ needed to match asteroseismic observations will decrease. In prior models, Tayler torques would often enforce $q \simeq q_{\rm min}$, but this is not the case in the updated prescription, so the rotation profiles are slightly different, but shear is still predicted to be largest where $N$ is largest. Upon implementing this AM transport model into the same stellar models as \cite{fuller:19}, we find that a value of $\alpha \sim 0.25$ produces very similar core rotation rates as the prior prescription, which had $\alpha \sim 1$ but only operated when $q > q_{\rm min}$.  Since we expect $\alpha$ to be of order unity, a value of $0.25$ is reasonable and suggests this model is viable. Hence we adopt $\alpha = 0.25$ as an asteroseismically calibrated saturation coefficient for the models in this paper. 

As discussed in Section 4 of \cite{ma:19}, the evolutionary time scale of the star is always longer than the Tayler instability growth time such that we can assume the instability has reached the statistical equilibrium discussed above. In some cases, however, the AM transport time scale can become longer than the stellar evolutionary time, such that Tayler torques become ineffective. In these cases, we expect AM to be nearly conserved in radiative regions, unless other processes such as wave-driven AM transport (e.g., \citealt{fullerwave:15}) are important.

\section{Stellar Evolution Models}
\label{stellar}

Using the AM transport prescription described above, we run binary stellar evolution models using MESA \citep{paxton:11,paxton:13,paxton:15,paxton:18,paxton:19}, version 10398. Our first step is to create a grid of helium star models to use in our binary evolution modeling. To do this, we begin with zero-age main sequence (ZAMS) stars of masses $M_{\rm ZAMS} = \{14,16,18,20,25,30,40,50,60,70,80,90 \} \, M_\odot$. All models are initialized with a rotation period of $P_{\rm rot} = 2 \, {\rm d}$ (corresponding to rotational velocity $v_{\rm rot} \sim 200 \, {\rm km/s}$) and have metallicity $Z=0.02$, except the $80 \, M_\odot$ and $90 \, M_\odot$ models which have $Z = 0.002$. As discussed in \cite{ma:19}, the final core AM is not very sensitive to the initial rotational velocity, so we do not vary the that parameter.

Throughout the evolution, wind mass loss is included via MESA's ``Dutch" wind prescription \citep{dejager:88,vink:00,nugis:00,vink:01} with efficiency $\eta = 0.5$. Moderate exponential cnvective overshoot is included using \verb|overshoot_f|$=0.02$. In convective zones, we adopt MESA's default AM transport in which convection acts like an AM viscosity of $\nu \sim \! H v_{\rm con}$, where $H$ and $v_{\rm con}$ are the scale height and convective velocity, such that rigid rotation is maintained. We also use MLT++ \citep{paxton:13} and set the surface to an optical depth of $10^3$ in order to bypass structural issues related to super-Eddington luminosities in the near-surface layers. This decreases the radius of the star by a tiny amount and is unlikely to affect the core evolution or AM content. MESA inlists are provided at this Zenodo repository.\footnote{https://zenodo.org/record/5778001\#.YdSDDmjMLb1}

We then evolve these stars in a binary with initial orbital period of $50 \, {\rm d}$ with a $1.5 \, M_\odot$ point-mass companion. This choice allows the stars to effectively evolve as single stars until they overflow their Roche lobes when crossing the Hertzsprung gap, after they have exhausted hydrogen in their core but before helium burning has begun. Upon Roche lobe overflow, we simulate mass transfer by stripping off the hydrogen envelope using MESA's \verb|relax_mass| feature. This results in helium stars of masses $M_{\rm He} = \{3.4,4.2,5.0,5.8,7.5,9.7,14,20,23,27.5,35,41 \} \, M_\odot$, which are the template models we use for further calculations.

For each helium star model, we then relax its metallicity to either $Z = 0.02$ or $Z=0.002$. While forming a $Z=0.002$ model from a $Z=0.02$ progenitor model (or vice versa) is not fully self-consistent, it allows us to explore a wider set of He star models. We shall see that the resulting spins depend primarily on final orbital period and final mass, with little memory of how the star got to that state (i.e., the initial mass, metallicity, companion mass, and orbital period). Next, we place the helium stars into an orbit with an equal-mass companion (treated as a point mass), with an orbital period of $P_{\rm orb} = \{0.25, 0.5, 0.75, 1.0, 1.25, 1.5, 2.0, 99 \} \, {\rm days}$. The period cannot be much smaller than $0.25 \, {\rm days}$ without causing Roche lobe overflow during the He-burning stage, while periods longer than 2 days are uninteresting because tidal torques become negligible according to our tidal prescription. The 99-day models are effectively single helium stars.

For tidal torques, we use the same method as \cite{qin:18} (see also \citealt{yoon:10,kushnir:17}) based on \cite{zahn:77}. For a star of mass $M$ and radius $R$, the tidal torque can be written
\begin{equation}
\dot{J}_{\rm tide} = 3 q^2 (\Omega_{\rm orb} - \Omega) M R^2 \left(\frac{R}{a}\right)^{\! 6} \frac{k}{T} \, ,
\end{equation}
where $\Omega_{\rm orb}$ is the angular orbital frequency, and $q = M'/M$ is the mass ratio of the companion. Here, $k/T$ describes how well the tidal forcing couples with internal gravity waves launched from the outer boundary of the convective core, which in Zahn's notation is 
\begin{equation}
\frac{k}{T} = \omega_{\rm d} (1 + q)^{5/6} \left(\frac{R}{a}\right)^{\! 5/2} E_2 \, ,
\end{equation}
where $\omega_{\rm d}= \sqrt{GM/R^3}$ is the star's dynamical frequency, and $E_2$ scales strongly with the core radius. To evaluate this factor, we follow \cite{kushnir:17} who show
\begin{equation}
\frac{k}{T} \simeq \frac{2}{3} \frac{\omega_{\rm d}^2}{\omega_{\rm c}} \left(\frac{r_{\rm c}}{R}\right)^{\! 5} \left(\frac{\omega_{\rm f}}{\omega_{\rm c}}\right)^{\! 5/3}\, \frac{\rho_{\rm c}}{\bar{\rho}_{\rm c}} \left(1 - \frac{\rho_{\rm c}}{\bar{\rho}_{\rm c}} \right)^2 \, ,
\end{equation}
with $\omega_{\rm c} = \sqrt{G m_{\rm c}/r_{\rm c}^3}$, $\omega_{\rm f} = 2 (\Omega_{\rm orb} - \Omega)$, with $m_{\rm c}$ and $r_{\rm c}$ the convective core mass and radius, while $\rho_{\rm c}$ is the density at $r_c$ and $\bar{\rho}_{\rm c}$ is the average core density.

We use the same implementation as used in \cite{qin:18}, in which the torque is applied uniformly per unit mass to the radiative envelope, such that the rate of change of specific AM of each fluid element is
\beq
\label{jdot}
\dot{j} = \frac{2}{3 \kappa} \frac{r^2}{R^2} \frac{\dot{J}_{\rm tide}}{M} \, ,
\eeq
and $\kappa$ is the gyration radius $\kappa = I/(MR^2)$, with $I$ the star's moment of inertia. Technically, $\kappa$ should be the gyration radius of the radiative envelope such that $\int dm \dot{j} = \dot{J}_{\rm tide}$, but in our models this is nearly equal to $\kappa$ so the distinction is unimportant. Another technical detail is that equation \ref{jdot} is only appropriate if $\Omega$ is constant within the envelope. Since our AM transport prescription predicts nearly rigid rotation during core He burning when the tidal torque is important, this issue is not problematic. The tidal torque vanishes when the convective core disappears after core He burning and resumes when a convective C-burning core or He-burning shell appears, but this typically occurs in the final hundreds of years at which point tidal torques become negligible. 

We then evolve these stars until the onset of CC. Some models halt due to numerical problems after oxygen burning, but we find that subsequent AM transport is negligible so the core AM content is very close to its value at CC. The $3.4 \, M_\odot$ models lose a large fraction of their helium envelopes via case BB mass transfer, and they undergo off-center neon burning, at which point their evolution is terminated. It is thus possible that their final core AM are slightly overestimated due to AM transport during Ne/O burning.

In subsequent sections, we will compute the time until merger due to gravitational wave emission after CC, in the absence of kicks. To compute the post-CC orbital semi-major axis, eccentricity, and orbital decay time, we follow the procedure described in \cite{tauris:17} (see also \citealt{blaauw:61}). For BH formation, this entails no change in the eccentricity or semi-major axis because we assume the entire star collapses into the BH. For NS formation, we assume that all the progenitor mass, apart from $1.6 \, M_\odot$, is instantaneously lost from the system.

\subsection{Angular Momentum Extraction}
\label{amev}

\begin{figure}
\includegraphics[scale=0.34]{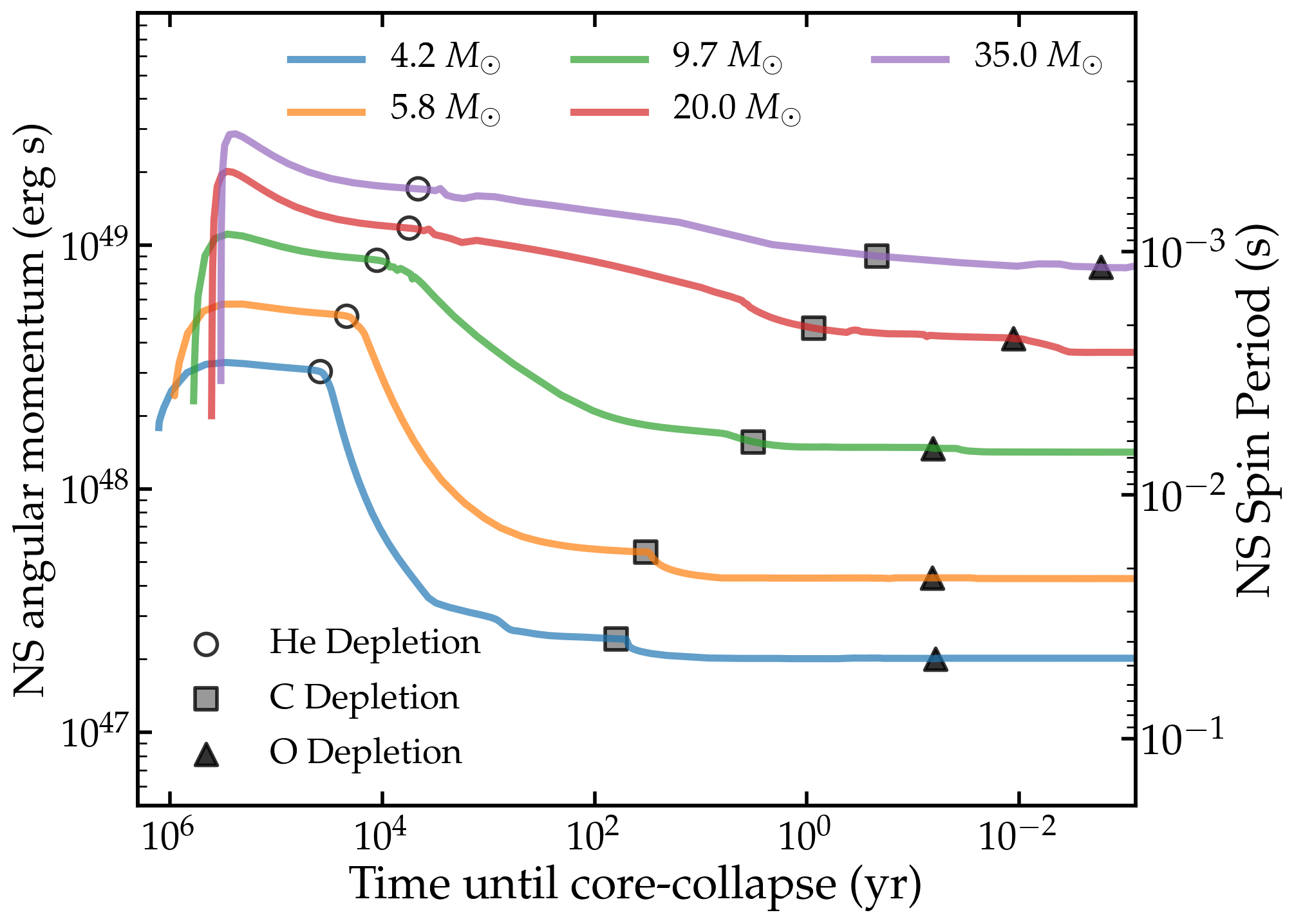}
\caption{\label{Jns} The angular momentum (AM) content of the inner $1.6 \, M_\odot$ of helium stars of various masses, as a function of time. The right axis shows the corresponding NS rotation period. Each of these models is initialized with an orbital period of $P_{\rm orb} = 0.5 \, {\rm day}$. The core AM initially increases due to tidal spin-up, but it subsequently decreases as the core contracts and AM is transferred to the outer layers. 
}
\end{figure}

The evolution of the star's core AM provides useful insight into the AM transport processes affecting the rotation rate of the compact object born upon CC.  \autoref{Jns} plots $J_{\rm core}$ (the AM contained within the inner $1.6 \, M_\odot$) of several different helium star models, each initialized with $Z=0.002$ at an orbital period of $P_{\rm orb} = 0.5 \, {\rm d}$. The mass cut of $1.6 \, M_\odot$ is a reasonable estimate for the baryonic mass of a NS born in the case of a subsequent explosion. Assuming conservation of AM during CC and explosion, the corresponding NS rotation period is $P_{\rm NS} = 2 \pi I_{\rm NS}/J_{\rm core}$. We use a NS moment of inertia of $I_{\rm NS} = 1.5 \times 10^{45} {\rm g} \, {\rm cm}^2$, reasonable for this mass \citep{worley:08}. 

The helium star's initial core AM is determined primarily by AM transport within the progenitor as it expands after the main sequence. As discussed in \citealt{ma:19}, most of the core's AM is lost during this phase, before the onset of envelope stripping. The core's AM is nearly conserved during the stripping process, so our models self-consistently determine the initial spin rate of the helium star based on the progenitor's evolution. In our models, the core's AM is remarkably insensitive to the rotation rate of the progenitor because the internal torques are proportional to $\Omega^4$, so that the core's spin reaches an ``equilibrium" spin rate at which spin-up via contraction is balanced by spin-down due to AM transport to the envelope.

At the short orbital period $P_{\rm orb} = 0.5 \, {\rm d}$ in \autoref{Jns}, tidal synchronization is fairly efficient and spins the stars up towards synchronous rotation, increasing $J_{\rm core}$. The most massive models in \autoref{Jns} achieve nearly synchronous rotation due to their larger core radii (and hence stronger tidal torques), whereas the least massive models are only partially synchronized by the end of core He-burning. Each model remains rigidly rotating during core He-burning. Due to mass loss via winds during this phase, the orbits widen slightly, decreasing the orbital/spin frequency. The core contracts slightly as He is depleted, reducing its moment of inertia and AM because it remains coupled with the radiative envelope. These effects combine to produce a slight reduction of $J_{\rm core}$ after tidal spin-up during core He-burning. 

\begin{figure}
\includegraphics[scale=0.36]{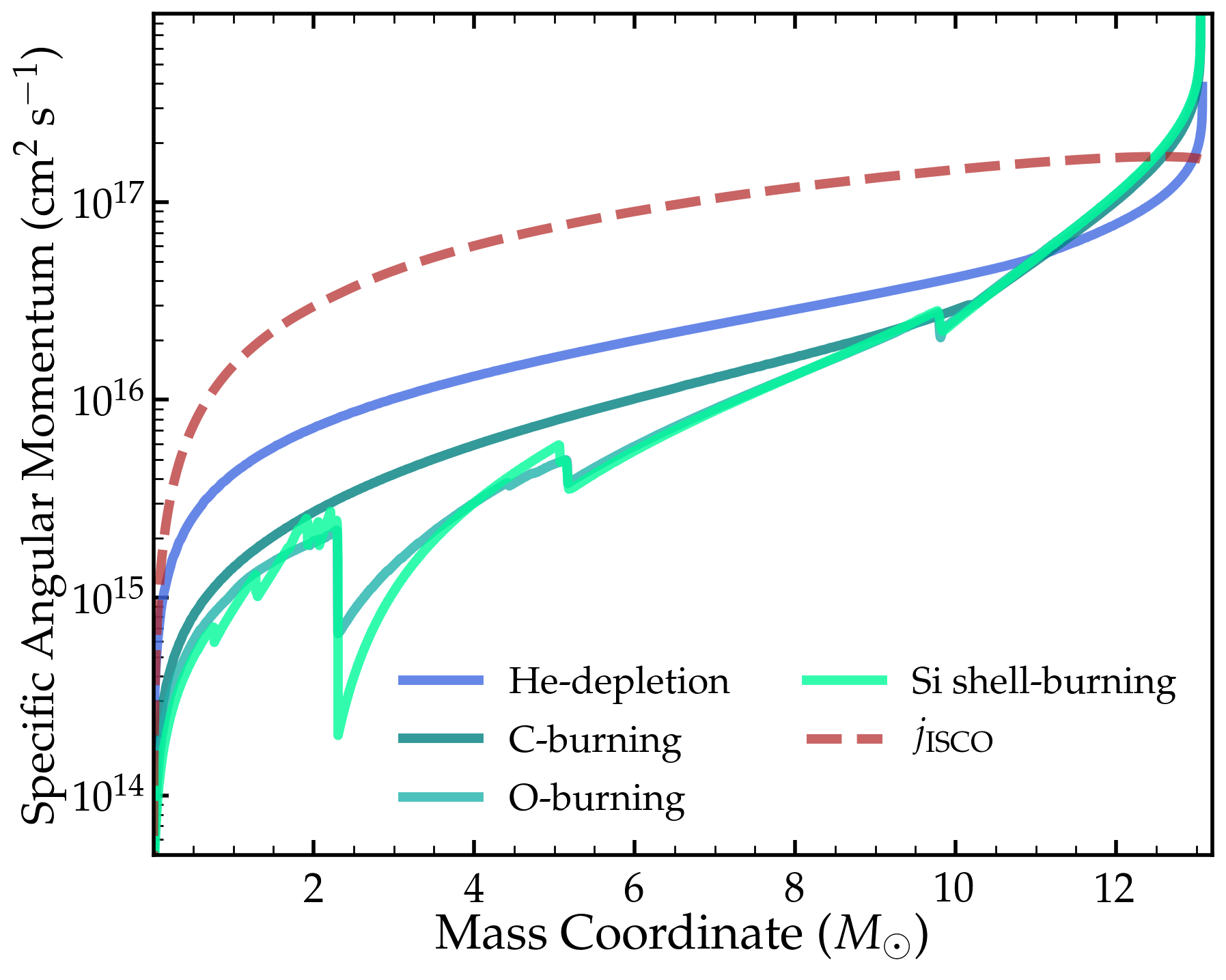}
\caption{\label{40jprof} Spherically averaged specific angular momentum as a function of mass coordinate within our $M_{\rm He} = 14 \, M_\odot$ ($M_{\rm ZAMS} = 40 \, M_\odot$), $Z=0.002$ helium star with orbital period $P_{\rm orb} = 0.5 \, {\rm d}$. Lighter lines correspond to progressively later evolutionary stages, while the red dashed line corresponds to the specific angular momentum needed to circularize outside a black hole formed from the interior layers. 
}
\end{figure}

The largest extraction of core AM occurs between core He-depletion (denoted with open circles) and core C-depletion (denoted with gray circles). During this phase, the core contracts greatly but AM transport remains efficient, keeping the star nearly rigidly rotating until the beginning of C-burning. The large decrease in the core's moment of inertia thus greatly decreases its AM content. During this phase, AM transport is efficient because the CO core has nearly uniform composition, hence there is not a strong composition gradient contributing to the compositional buoyancy frequency $N_\mu$ that restricts AM transport (see \citealt{fuller:19} for more detailed discussion). 

During and after C-burning, large amounts of differential rotation begin to appear within the CO core, due to the large composition gradients that develop, and the shorter evolutionary times. However, the short remaining lifetime of the star means that only a small amount of AM can subsequently be extracted from the core before it collapses. Hence, $J_{\rm core}$ asymptotes to its final value soon after C-depletion, in spite of the fact that the rate at which the core loses AM generally continues to increase. We find that the AM extraction rates after He-depletion are similar between the low and high-mass models. However, the larger initial values of $J_{\rm core}$ and the much shorter evolution time scales of the high-mass models means that much more core AM is retained at CC. Consequently, a NS produced from a $4.2 \, M_\odot$ model is predicted to rotate with a period $P_{\rm NS} \! \sim \! 50 \, {\rm ms}$, while a NS produced from a $35 \, M_\odot$ model is predicted to rotate with a period $P_{\rm NS} \! \sim \! 1 \, {\rm ms}$.

\autoref{40jprof} shows the specific AM as a function of mass coordinate for a $M_{\rm He} = 14 \, M_\odot$ model in a $P_{\rm orb} = 0.5 \, {\rm d}$ binary, plotted at a few different evolutionary stages. This model has a $\sim \! 3 \, M_\odot$ helium envelope and a radius of $\sim \! 1.2 \, R_\odot$ at core-collapse. It would produce a NS with $P_{\rm NS} \! \approx \! 4 \, {\rm ms}$ or a BH with $a_{\rm BH} \! \approx \! 0.5$ (see Section \ref{rotationrate}). The specific AM of the core decreases by a factor of a few between central helium depletion and central carbon burning, and decreases only slightly more before CC. Note that the specific AM of the outermost layers actually increases due to the AM transported outward from inner layers. Only the outer $\sim \! 1 \, M_\odot$ has enough AM ($j > j_{\rm ISCO}$) to circularize outside of a BH formed from the interior layers (see Section \ref{engine} for calculation of $j_{\rm ISCO}$).

\section{Compact Object Rotation Rates}
\label{rotationrate}

\subsection{Neutron Star Rotation Rates}

\begin{figure}
\includegraphics[scale=0.33]{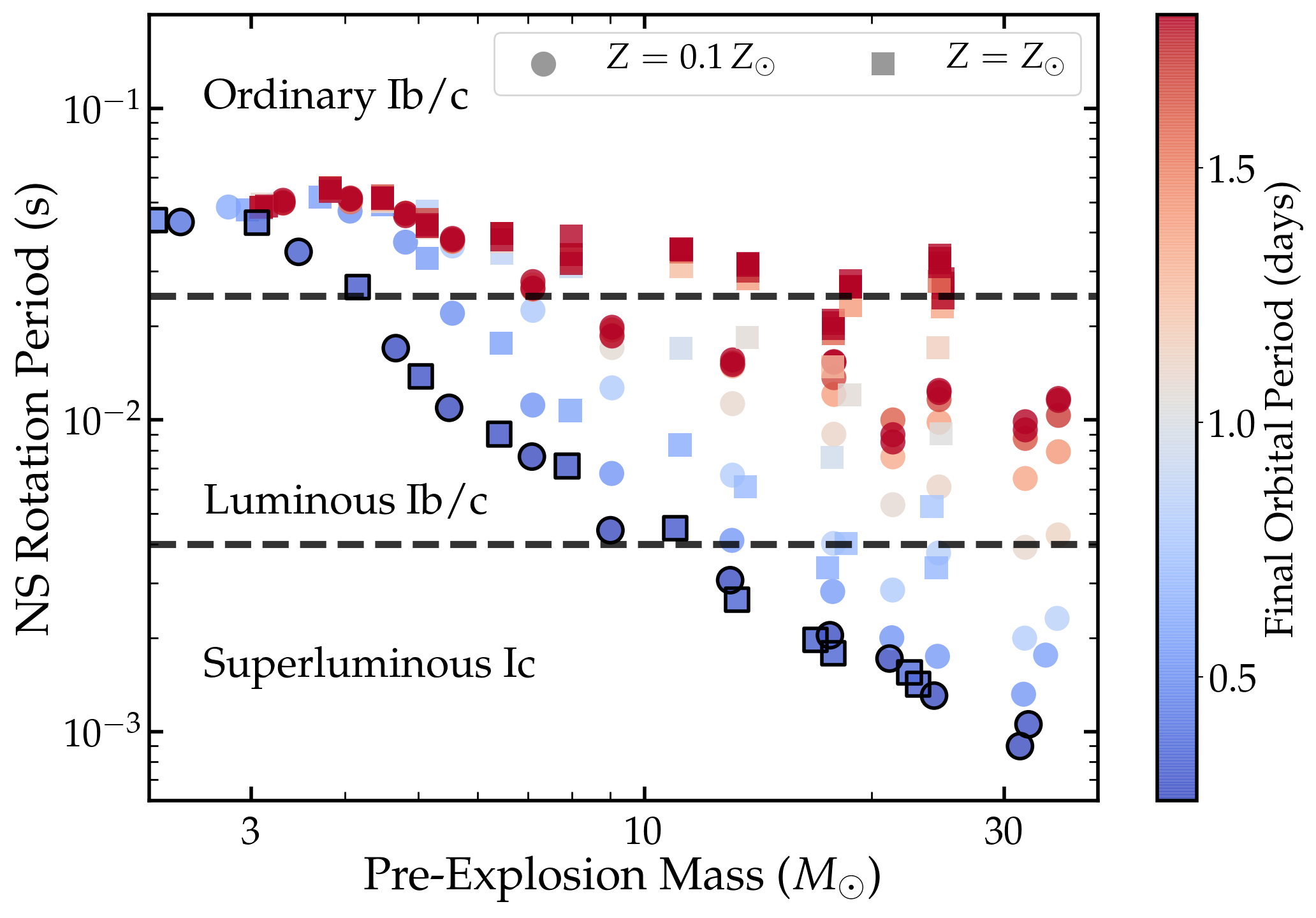}
\caption{\label{Pns} The natal neutron star rotation period for each of our models, calculated from the core angular momentum, as a function of the pre-explosion mass of its helium progenitor star. The color of each symbol indicates the binary orbital period at core-collapse. Squares and circles denote solar metallicity and low-metallicity models, respectively. Black outlines denote models that would merge within a Hubble time. Dashed lines suggest the approximate boundaries between ordinary, luminous, and superluminous supernovae, given an appropriate magnetar field strength.
}
\end{figure}

\autoref{Pns} shows our predictions for the natal NS rotation rates of each of our models, assuming 1) a successful explosion, 2) a NS of baryonic mass $1.6 \, M_\odot$ and moment of inertia $I_{\rm NS} = 1.5 \times 10^{45} {\rm g} \, {\rm cm}^2$, and 3) conservation of AM during CC and explosion. Our models predict a strong correlation between the pre-explosion mass and the NS rotation period, with higher mass stars producing NSs that rotate up to 50 times faster than low-mass stars, down to rotation periods of $P_{\rm NS} \sim 1 \, {\rm ms}$ for the range of models considered here. While many of these stars likely produce BHs rather than NSs upon CC, those NSs that are born from massive stars in tight binaries are likely to be much more rapidly rotating than typical young NSs.

There is also a strong correlation between pre-explosion orbital period and the NS rotation period, with rapidly rotating NSs arising from more compact binaries due to the more rapid rotation enforced by tidal synchronization during core He-burning. For stars in tight binaries, the progenitor metallicity only indirectly affects the NS rotation period, due to its affect on stellar winds which decrease $M_{\rm exp}$ and increase $P_{\rm orb}$ due to mass loss during the He-burning phase. This makes rapid NS rotation more likely in low-metallicity environments, as also found for BH spin (e.g., \citealt{qin:18,bavera:19}). Because $P_{\rm orb} \lesssim 1 {\rm d}$ is required for millisecond NS periods in this scenario, many massive ($M \gtrsim 5 \, M_\odot$) main sequence companions are too large to fit within the orbit, and tidal spin-up is more likely to be achieved by a compact object or a low-mass companion star. 

At orbital periods $P_{\rm orb} \gtrsim 1 \, {\rm d}$, tidal synchronization does not occur and the NS rotation rate will depend on the progenitor rotation rate. We find relatively long NS rotation periods ($P_{\rm NS} \gtrsim 10 \, {\rm ms}$) for all the wide binary models considered here.
Slower NS rotation may be possible for stars born less rapidly rotating or with higher mass loss rates due to higher metallicity. Faster NS rotation may be possible for stars born more rapidly rotating or with lower mass loss rates due to lower metallicity.

For NSs originating from low-mass progenitors with $M_{\rm He} \lesssim 4 \, M_\odot$, the predicted NS rotation rate is $P_{\rm NS} \sim 50 \, {\rm ms}$ with little dependence on mass and orbital period. This feature is similar to that found by \cite{fuller:19} in which the predicted rotation rates of WDs are only weakly dependent on the progenitor's rotation rate. The longer evolution time scales in these stars allow the torques arising from Tayler instability to achieve ``equilibrium" in the sense that they reduce the core rotation rate until the AM transport time scale (which scales as $\Omega^{-3}$) becomes longer than the stellar evolution time scale. Evidently this ``equilibrium" state is not reached in the advanced evolution of high-mass He stars because of their very short evolution time scales, leading to a spread in NS rotation rates that depend on the progenitor rotation rate. 

If NS baryonic masses from very massive stars are larger than $1.6 \, M_\odot$, they will contain more AM and could spin faster than the models in \autoref{Pns}. We find that using a baryonic mass of $2.3 \, M_\odot$ and assuming $I_{\rm NS} \propto M_{\rm NS}$ results in spin periods that are smaller by a factor of $\sim \! 1.5$. Smaller NS masses result in only slightly larger NS spin periods. Hence, the possible range of NS periods can only vary by a factor of $\sim \! 2$ due to differences in the NS mass. 

We note that the NS rotation periods of our single He stars models are a factor of $\sim 4$ shorter than those born form H-rich stars in \cite{ma:19}, when comparing to the same ZAMS progenitor mass. There are two reasons. The first is that the helium core in H-rich stars is embedded in a H envelope that continues to extract AM from the He core, producing slightly slower core rotation rates. The second is that the updated AM transport prescription discussed in \autoref{angmom} leads to rotation rates faster by a factor of $\sim 2$ compared to the prescription of \cite{ma:19}. As both of these prescriptions produce similar results for low-mass stars which approximately match asteroseismic data, this difference reflects the uncertainty of the models when extended to massive stars.

The NS rotation rates of our models can be reasonably approximated by a power-law relationship with the pre-explosion mass and orbital period, for pre-explosion masses $M_{\rm exp} \gtrsim 5 \, M_\odot$ and $P_{\rm orb} \lesssim 1 \, {\rm days}$. We find that
\begin{equation}
P_{\rm NS} = 1.4 {\rm ms} \left( \frac{M_{\rm exp}}{30 M_\odot} \right)^{\! -1.6} \left( \frac{P_{\rm orb}}{0.5 \, {\rm d}} \right)^{\! 1.0}
\end{equation}
provides a good fit to most of these models, with scatter at the level of $\sim \! 50 \%$. Given that uncertainties in the physics of AM transport are larger than this scatter, this approximation is good enough to be used in binary population synthesis calculations. For low-mass progenitors with $M_{\rm He} \lesssim 4 \, M_\odot$, a good approximation is simply $P_{\rm NS} \approx 50 \, {\rm ms}$. 

For He stars that are in binaries with $P_{\rm orb} \gtrsim 1 \, {\rm day}$, we find that 
\begin{equation}
P_{\rm NS} = 25 {\rm ms} \left( \frac{M_{\rm exp}}{30 M_\odot} \right)^{\! -0.25}
\end{equation}
provides a reasonable fit to our high-metallicity models, while
\begin{equation}
P_{\rm NS} = 10 {\rm ms} \left( \frac{M_{\rm exp}}{30 M_\odot} \right)^{\! -0.6} 
\end{equation}
provides a reasonable fit to our low metallicity models. It is possible that the star's initial rotation rate (which was not varied between our models) and mass loss prescription could affect the NS rotation period in this case, especially for high-mass stars. The dependence on metallicity for single He stars reflects the difference in rotation rate due to AM loss via stellar winds: low metallicity stars lose less mass/AM via winds and remain more rapidly rotating, as discussed in many prior works (e.g., \citealt{georgy:13}). However, the dependence on initial rotation rate and metallicity is weaker in our models due to the tendency for AM transport to push the core spin rate towards an equilibrium value as discussed in \autoref{amev} and \autoref{modeling}.

\subsection{Black Hole Spins}

\begin{figure}
\includegraphics[scale=0.33]{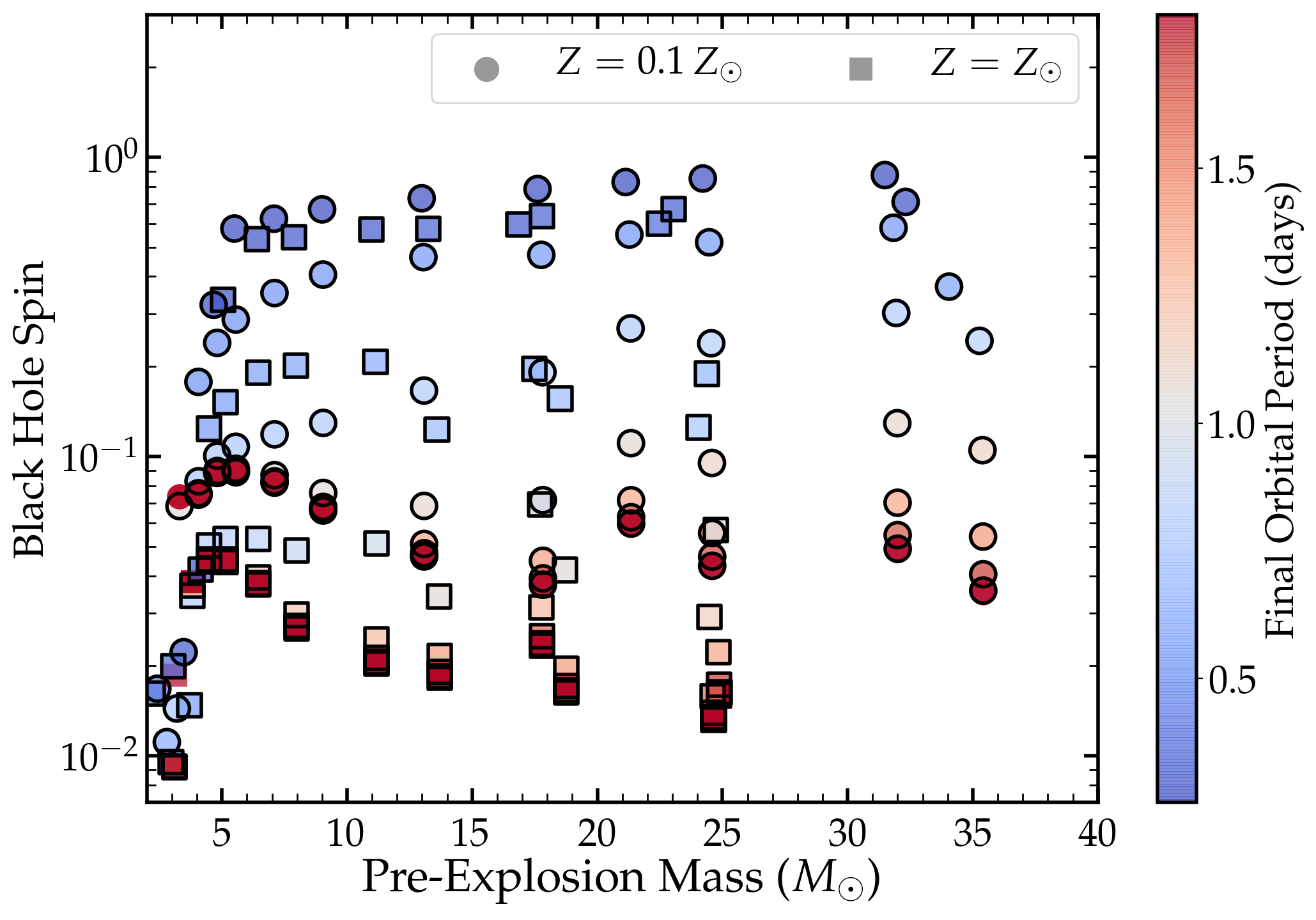}
\caption{\label{abh} The same as Figure \ref{Pns}, but showing the corresponding dimensionless black hole spin $a_{\rm BH}$, computed via equation \ref{abh2}.
}
\end{figure}

To estimate the spins of BHs formed from our models, we first compute the specific AM corresponding to the innermost stable circular orbit (ISCO)
\begin{equation}
j_{\rm ISCO} = \frac{2}{3^{3/2}} \left(1 + 2 \sqrt{3 r_{\rm ISCO}-2} \right) \frac{G m}{c}
\end{equation}
where $m$ is the interior mass and the ISCO radius $r_{\rm ISCO}$ is given by 
\begin{equation}
r_{\rm ISCO} = 3 + z_2 - \sqrt{(3-z_1)(3 + z_1 + 2 z_2)} \, ,
\end{equation}
with $r_{\rm ISCO}$ in units of $G m/c^2$. Here
\begin{equation}
z_1 = 1 + (1-a^2)^{1/3}\left( (1+a)^{1/3} + (1-a)^{1/3} \right) \, ,
\end{equation}
$z_2 = \sqrt{3 a^2 +z_1^2}$, and $a = J_{\rm ac}(m) c/(G m^2)$ is the spin of the BH with mass $m$ and AM $J_{\rm ac}(m)$. For each infalling shell in the progenitor, we assume that all the underlying mass has fallen into the BH. However, for any mass with with specific AM $j > j_{\rm ISCO}$, we assume that the corresponding accreted specific AM is $j_{\rm ISCO}$. Hence, the accreted AM is 
\begin{equation}
J_{\rm ac}(m) = \int_0^m {\rm min} (j,j_{\rm ISCO}) \, dm \, .
\end{equation}
For $j > j_{\rm ISCO}$, this crudely models accretion through a disk in which the accreted mass has the specific AM corresponding to the inner edge of the disk. The BH spin is then 
\begin{equation}
\label{abh2}
a_{\rm BH} = \frac{J_{\rm ac}(M) c}{G M^2} \, ,
\end{equation}
where $M$ is the mass of the pre-collapse progenitor.

Only a small amount of mass is lost due to neutrino emission from He stars \citep{fernandez:18}, but there could be additional loss of BH AM due to material ejected during a failed explosion or via accretion feedback \citep{batta:19}. Conversely, the BH spin could be increased after accretion of fallback material that gains AM due to interaction with the companion \citep{batta:17,schroder:18}. In our simple accretion disk modeling of Section \ref{engine} (see Appendix \ref{accretion}), most of the material that circularizes outside of ISCO is blown away by disk winds, slightly decreasing the BH mass and typically decreasing $a_{\rm BH}$ by $\sim \! 0.2$ for high-spin models. Hence, our method above is likely to slightly overestimate BH spins. 

\autoref{abh} shows our predictions for BH spins as a function of progenitor mass and orbital period. We see that the predicted BH spin is nearly independent of progenitor mass or metallicity for $M_{\rm exp} \gtrsim 5 \, M_\odot$ and $P_{\rm orb} \lesssim 1 \, {\rm d}$ for which tidal synchronization occurs. In this regime, the spin is determined by the orbital period, with $a_{\rm BH}$ near unity for models with final orbital periods $P_{\rm orb} \lesssim 0.5 \, {\rm days}$. This is because the progenitor's spin frequency is approximately equal to the orbital frequency due to tidal synchronization at short orbital periods. An exception occurs for the lowest values of $M_{\rm exp}$, which have much smaller values of $a_{\rm BH}$. These stars have much smaller moments of inertia due to their more centrally concentrated structures, so that their AM content $J$ is small even if they are tidally spun up. In most cases, however, we expect these stars to produce NSs rather than BHs.

While there is little explicit dependence of $a_{\rm BH}$ on metallicity when expressed in terms of the final orbital period, higher metallicity systems will lose more mass during core He-burning and hence evolve to longer orbital periods, as discussed above and in \cite{qin:18} and \cite{bavera:19}. This can be seen in \autoref{abh}, as the low-metallicity models have systematically larger values of $M_{\rm exp}$ and $a_{\rm BH}$ even though they are initialized at the same helium star mass and with the same orbital periods. Hence, we expect higher spins in low-metallicity environments where there is less orbital widening.

Because of their large masses, most of the BH systems shown in \autoref{abh} will merge within a Hubble time. Assuming equal mass BHs in a circular binary, the maximum orbital period for which a merger will occur within $t_{\rm GW} = 13 \, {\rm Gyr}$ is 
\begin{equation}
\label{pmax}
P_{\rm max} \simeq 2.3 \, {\rm d} \left( \frac{M_{\rm BH}}{10 M_\odot} \right)^{\!5/8} \, . 
\end{equation}
Hence, all of the rapidly rotating BHs with $P_{\rm orb} \lesssim  1 \, {\rm d}$ are expected to merge in much less than a Hubble time, while many slowly rotating BHs with $P_{\rm orb} \gtrsim 1 \, {\rm d}$ are also expected to merge within a Hubble time. The observed spins of BHs are strongly dependent on the common envelope efficiency that sets their orbital period (or alternatively, the amount of orbital decay in the stable mass transfer scenario), with a more moderate dependence on progenitor metallicity. 

Similar to the NS scenario discussed above, the short orbital periods required for high BH spins preclude many massive main sequence stars as the companion stars to tidally spun-up helium stars. Hence, the second BH can be more easily formed with high spin in this binary scenario, and we expect the first BH to have a low spin similar to the long-period systems in \autoref{abh}. Therefore, the expected $\chi_{\rm eff}$ measured by LIGO will likely be $\chi_{\rm eff} \lesssim 0.5$ in most cases, assuming the second formed BH has lower mass than the first formed BH.

The values of $a_{\rm BH}$ of our models with $P_{\rm orb} \lesssim 1 \, {\rm d}$ and $M_{\rm exp} \gtrsim 5 \, M_\odot$ can be fairly well approximated by a simple power-law fit,
\begin{equation}
a_{\rm BH} = 1 \left( \frac{M_{\rm exp}}{30 M_\odot} \right)^{\! 0.5} \left( \frac{P_{\rm orb}}{0.43 \, {\rm d}} \right)^{\! -2.5} \, .
\end{equation}
Similar to our results for NSs, the scatter about these approximations is typically less than $\sim 50\%$ such that they are good enough to be used in population synthesis calculations. See \cite{bavera:21} for a more complicated (but more accurate) fitting function.

For BHs born from He stars in single or wide binaries with $P_{\rm orb} \gtrsim 1 \, {\rm d}$, 
we find fitting functions of 
\begin{equation}
a_{\rm BH} = 0.013 \left( \frac{M_{\rm exp}}{30 M_\odot} \right)^{\! -0.6} \, 
\end{equation}
for solar metallicity ($Z=0.02$) and 
\begin{equation}
a_{\rm BH} = 0.042 \left( \frac{M_{\rm exp}}{30 M_\odot}\right)^{\! -0.3} \, 
\end{equation}
for low metallicity ($Z=0.002$). As with NSs, the BH spin in this case depends on metallicity due to AM losses via winds during He-burning, as well as the initial rotation rate of the progenitor star. However, our results suggest that BHs born in wide binaries will generally be slowly rotating with $a_{\rm BH} \lesssim 0.1$. A possible exception is for BHs born via the chemically homogeneous evolution channel, which may be rapidly rotating at very low metallicity for single stars (e.g., \citealt{yoon:06}) or stars in slightly wider binaries \citep{demink:16,mandel:16,marchant:17}. 

The BH spins here are a few times faster than those of H-rich stars from \cite{fullerma:19} because of the updated AM transport prescription and the continued loss of core AM to the envelope in H-rich stars. Our results for BH spin from He stars in tight binaries appear similar to those of \cite{qin:18}, \cite{bavera:19}, and \cite{bavera:21}, when tidal spin-up occurs. We agree that $a_{\rm BH}$ can approach unity for stars with final orbital periods $P_{\rm orb} \lesssim 0.4 \, {\rm d}$, and that $a_{\rm BH} \lesssim 0.1$ for final orbital periods $P \gtrsim 1 \, {\rm d}$, with moderate spins at periods between these values. Like \cite{bavera:21}, we find smaller spins for low-mass BHs due to the smaller progenitor moment of inertia during core He-burning, with a sharp falloff in spin for BH masses below $\sim \! 5 \, M_\odot$.
Similar to \cite{bavera:21}, BHs born from our most massive models ($M_{\rm He} \gtrsim 35 \, M_\odot$) have smaller spin, because those stars typically lose so much mass that their orbital period widens and tidal spin-up becomes ineffective.

Overall, the differing AM transport prescription between our work and models employing the TS dynamo has little effect on BH spins from He stars in compact binaries, which are primarily determined by tidal spin-up. The difference may be larger for stars in wider binaries where the He star's spin is primarily determined by AM transport within the hydrogen-rich progenitor, or when the He star is formed via chemically homogeneous evolution (see \autoref{homogeneous}).

\section{Energetic supernovae and transients}

Central engines are suspected to power energetic transients such as GRBs, type Ic-BL SNe, and superluminous type Ic SNe. Here, we examine the ability of our models to power these events through either accretion onto a BH after CC, or via the spin-down energy of a rapidly rotating magnetar. We also compare to observational inferences of magnetar rotation periods obtained by modeling type Ibc SNe. 

\subsection{Comparison with observed supernovae}
\label{observed}

\begin{figure}
\includegraphics[scale=0.33]{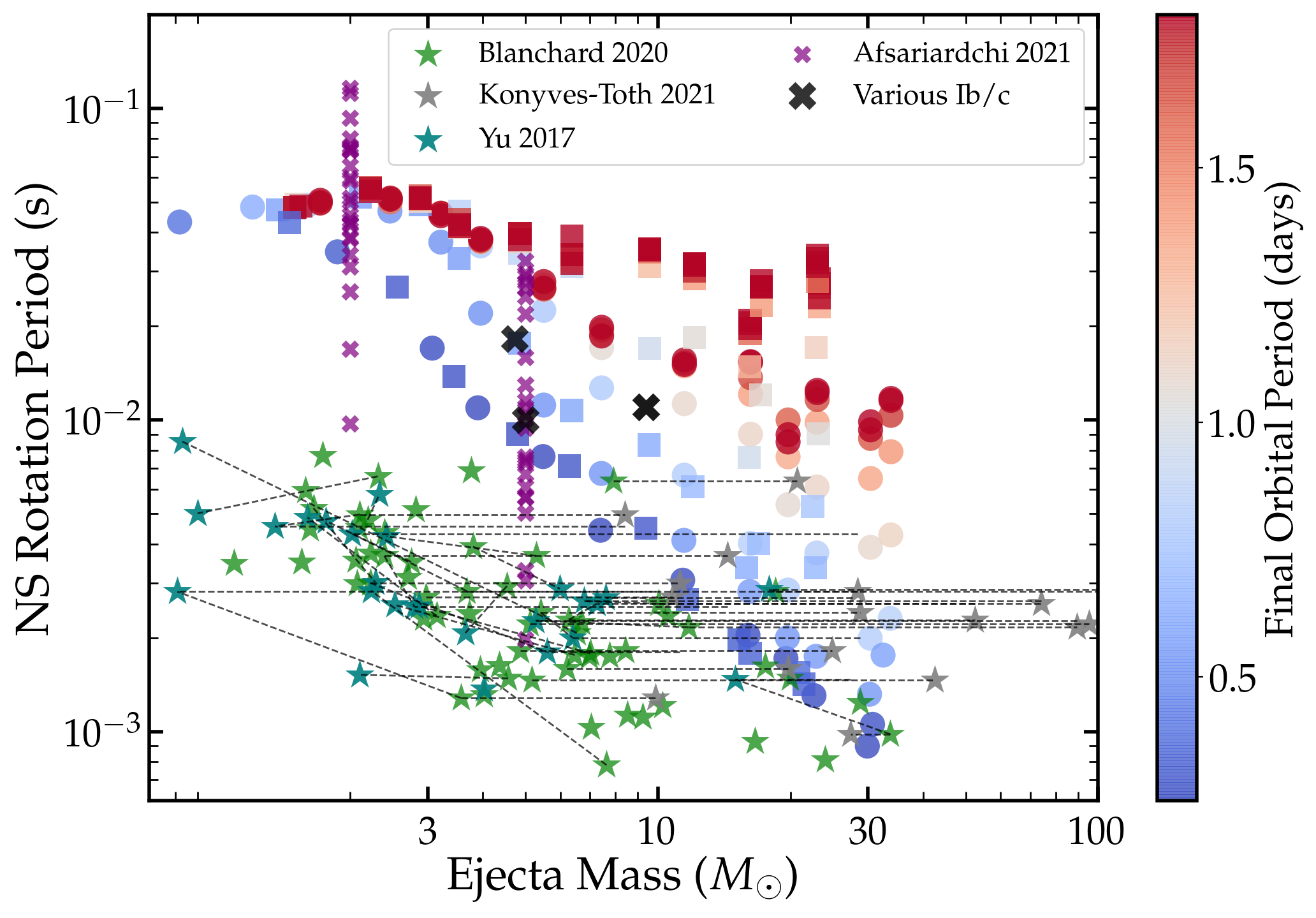}
\caption{\label{Pnsmeasured} The same as Figure \ref{Pns}, but now in comparison with inferred magnetar rotation periods and ejecta masses for superluminous type Ic SNe taken from \citealt{yu:17} and \citealt{blanchard:20}. When available, dashed lines connect different estimates for the same events, or with ejecta masses from \citealt{konyves-toth:21}. We also indicate possible magnetar spin periods from \citealt{afsariardchi:21}, where the ejecta mass has been assumed to be $2 \, M_\odot$ or $5 \, M_\odot$. Black crosses correspond to a few individually modeled type Ib/c SNe discussed in the text.
}
\end{figure}

\autoref{Pnsmeasured} compares our predicted NS rotation periods (the same as shown in \autoref{Pns}) to observationally inferred magnetar rotation periods, as a function of ejecta mass. \cite{yu:17} and \cite{blanchard:20} estimated ejecta masses, magnetar rotation periods, and magnetar field strengths by modeling SLSNe Ic light curves assuming they are powered by dipole spin-down of a rapidly rotating magnetar. Their inferred rotation periods and ejecta masses (teal/green stars in \autoref{Pnsmeasured}) show a similar trend to our predictions, with higher ejecta masses corresponding to more rapidly rotating magnetars. However, their inferred ejecta masses are several times lower than our predictions, for a given NS rotation period. Alternatively, their inferred spin periods are several times shorter than our predictions for a given ejecta mass, especially at low ejecta masses. One possibility for this discrepancy is that our model predicts too much AM extraction from massive stellar cores, and that NSs are born rotating several times faster than our predictions. 

Another possibility is that the inferred ejecta masses of \cite{yu:17} and \cite{blanchard:20} are systematically too small. They used semi-analytic light curve models with a constant opacity, assuming a magnetar central engine. \cite{konyves-toth:21} analyzed many of the same events and estimated ejecta masses using Arnett's rule (\citealt{arnett:80}, see also \citealt{khatami:19}), remaining agnostic to the source of centralized heating. Their models also assumed a constant opacity and were coupled with spectroscopic ejecta velocity estimates. As shown in \autoref{Pnsmeasured}, their ejecta mass estimates are typically several times larger and are usually closer to the predictions of our models, though \cite{konyves-toth:21} did not estimate magnetar spin periods. In both cases, the use of a constant and gray opacity could lead to systematic errors when inferring the ejecta mass. Furthermore, the true ejecta masses could be larger than the measured ejecta masses in non-spherically symmetric SNe in which a massive but dim component of the ejecta is masked by a bright, low-mass, higher velocity component of the ejecta. It is safe to say that the ejecta masses are very uncertain, and we encourage more detailed photometric and spectroscopic modeling of these events to refine ejecta mass measurements. 

\autoref{Pnsmeasured} also shows a few magnetar rotation period estimates obtained by modeling ``normal" type Ib SNe,  which likely arise from He stars formed in binaries (e.g., \citealt{zapartas:21}). \cite{afsariardchi:21} provided estimates for many events by assuming a typical ejecta mass of either $2 \, M_\odot$ or $5 \, M_\odot$. They found magnetar rotation periods very similar to those predicted by our models were consistent with helping to power the events they analyzed. \cite{ertl:20} came to a similar conclusion, showing that magnetars with periods of $P_{\rm ns} \sim 20 \, {\rm ms}$ may help power the luminous end of type Ibc SNe. Hence, the faint end of the type Ibc distribution may arise from helium star progenitors in wide binaries that produce slower rotating NSs/magnetars, while the bright end could be powered by moderately rotating magnetars in tight binaries.

The inferred ejecta masses and magnetar rotation periods for a few detailed models of individual type Ib/c SNe are also shown in \autoref{Pnsmeasured}. \cite{maeda:07} modeled the unusual type Ib SN 2005bf and concluded it could be powered by a magnetar with $P_{\rm NS} \sim 10 \, {\rm ms}$ and came from a progenitor with $M_{\rm ZAMS} \sim 20-25 \, M_\odot$, so we inferred an ejecta mass of $M_{\rm ej} \sim 5 \, M_\odot$ based on our models for that mass range. \cite{pandey:21} modeled the luminous type Ib SN 2012au and found that an ejecta mass of $\approx \! 5 \, M_\odot$ and magnetar with $P_{\rm NS} \approx 20 \, {\rm ms}$ could reproduce the data. \cite{gutierrez:21} modeled the double-peaked type Ic SN 2019cad and estimated a pre-explosion mass of $11 \, M_\odot$ ($M_{\rm ej} \sim 9.4 \, M_\odot$) and a magnetar rotation period of $P_{\rm NS} \sim 11 \, {\rm ms}$. Each of these events are consistent with our models for a final orbital period of $P_{\rm orb} \sim 0.5-1 \, {\rm day}$.

\subsection{Predictions for central engine models}
\label{engine}

\subsubsection{Magnetar Central Engines}

\begin{figure}
\includegraphics[scale=0.33]{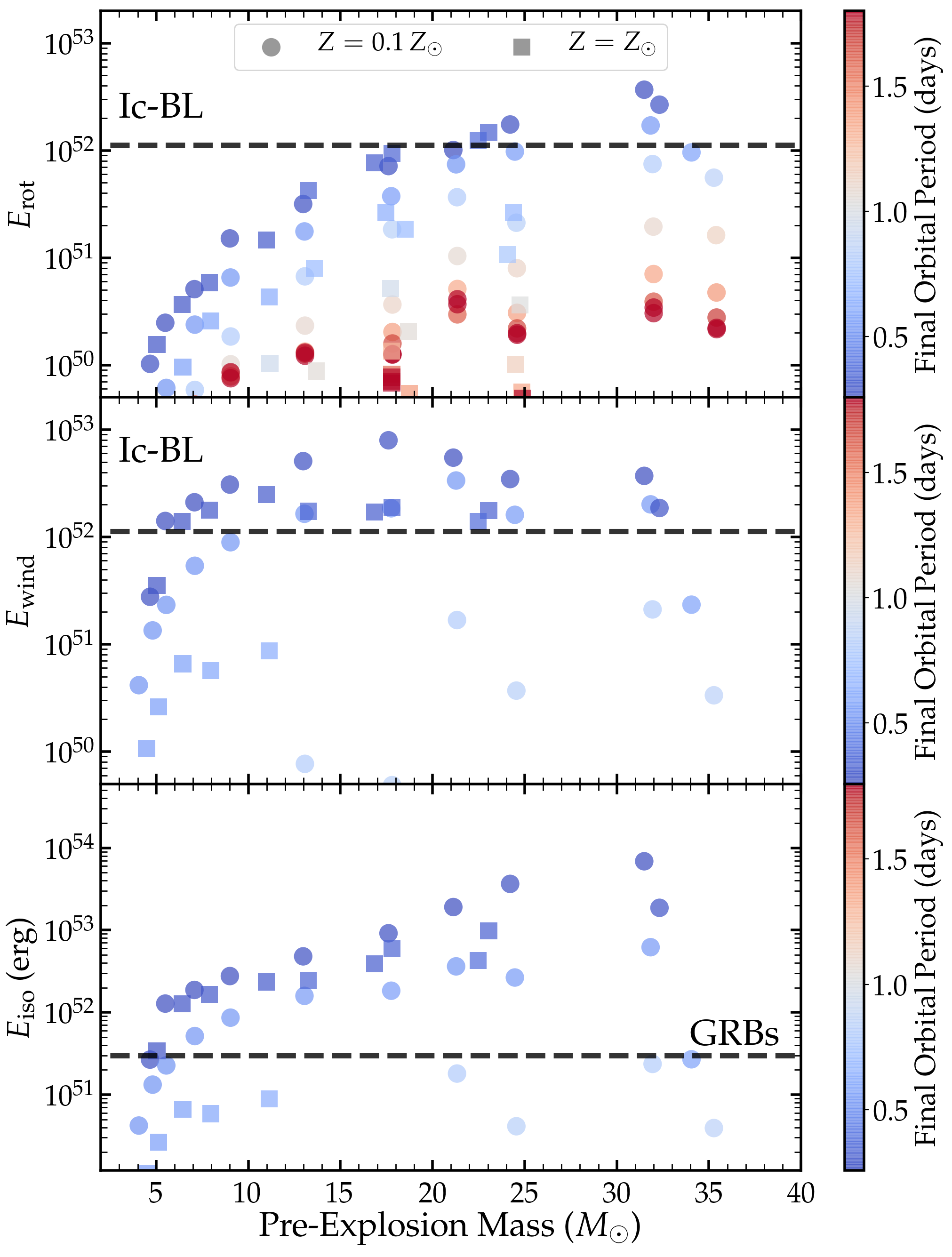}
\caption{\label{Eengine} Energetics of our models as a function of pre-explosion mass, with the same symbols as Figure \ref{Pns}. {\bf Top:} Isotropic GRB energy $E_{\rm iso}$ associated with accretion onto a newly formed BH for each of our models, as described in the text. Models with $E_{\rm iso} \gtrsim 3 \times 10^{51} \, {\rm erg}$ are good candidates to produce detectable GRBs. {\bf Middle:} Energy released from an accretion disk wind in the case of BH formation. {\bf Bottom:} Rotational energy $E_{\rm rot}$ upon formation of a $1.6 \, M_\odot$ NS. Rotational or disk wind energies larger than the dashed line may be capable of powering a Ic-BL SN via magnetar spin-down or BH accretion, respectively. Our models struggle to power Ic-BL SNe via magnetar spin-down, unless there is significant magnetar spin-up due to fallback accretion (see text).
}
\end{figure}

Our models also have implications for the progenitors of GRBs and type Ic-BL SNe. Many recent works \citep[e.g.,][]{metzger:11,mazzali:14,metzger:15,beniamini:17,margalit:18,barnes:18,fryer:19,aloy:21,shankar:21} have suggested rapidly rotating magnetars as the engines driving both of these events. To investigate this possibility, we compute the rotational energy $E_{\rm rot} = 0.5 I_{\rm NS} (2 \pi/P_{\rm NS})^2$ of a putative magnetar formed from the inner $1.6 \, M_\odot$ of each of our models.
We compare this to the typical energy $E_{\rm Ic-BL} \approx 0.5 M_{\rm ej} v_{\rm ej}^2 \sim 10^{52} \, {\rm erg} \,  (M_{\rm ej}/5 M_\odot) (v_{\rm ej}/1.5 \times 10^4 {\rm km/s})^2$ associated with observed Ic-BL SNe. The bottom panel of \autoref{Eengine} shows that the magnetar rotational energy of our models is less than $10^{52}$ erg for all but our most rapidly rotating models arising from the most massive progenitors. The discrepancy is even worse when one considers that the ejecta mass of our massive models would be far larger than $5 \, M_\odot$, meaning that the ejecta velocity would be smaller than $15,000 \, {\rm km/s}$ and those SNe would not appear broad-lined. Another problem with this scenario is that such massive He stars may be less likely to form via a common envelope event \citep{klencki:20,marchant:21,vanson:21}, and less likely to explode \citep{oconnor:11,zapartas:21}. This may disfavor magnetars as the power source of most Ic-BL SNe.

There are two possibilities that may reconcile our models with the magnetar hypothesis. The first is that Ic-BL explosions are highly aspherical (e.g., bipolar jet-driven explosions, \citealt{maeda:02}) such that a small fraction of mass absorbs most of the spin-down energy. This would drive larger ejecta velocities even for rotational energies less than $\sim \! 10^{52} \, {\rm erg}$.
The second possibility is that the magnetar is driven to large rotation rates by fallback accretion from outer layers of the star which lave larger specific AM. \cite{piro:11} (see also \citealt{metzger:18}) showed that this can temporarily drive magnetars to $\sim \! 1 \, {\rm ms}$ rotation periods (rotational energies $E_{\rm rot} \gtrsim 10^{52} \, {\rm erg}$), even if the magnetar initially rotates much more slowly. After fallback accretion has slowed, the magnetar would subsequently spin down via the propellor mechanism and/or dipole radiation, driving a type Ic-BL SN. In this scenario, Ic-BL can be produced by events with large magnetic fields of $B \sim 10^{15} \, {\rm G}$ and short fallback times of $t_{\rm fb} \sim 100 \, {\rm s}$, while SLSNe Ic can be produced by events with lower magnetic fields of $B\sim 10^{14} \, {\rm G}$ and long fallback times of $t_{\rm fb} \sim 10^4 \, {\rm s}$ \citep{lin:20,lin:21}.

Inspection of our models indicate that they do contain enough AM in their outer cores to spin-up a magnetar to $P_{\rm NS} \sim \! 1 \, {\rm ms}$. For instance, this is easily achievable in many of our models if $1 \, M_\odot$ of the inner $\sim 5 \, M_\odot$ falls back onto a NS made from the iron core. Still, it is not clear how often that amount of fallback can occur, whether that mass can accrete onto the magnetar before being blown away by a disk wind, nor how often it can drive the magnetar to rapid rotation without making it collapse to a BH. Further investigation of this possibility for our models is deferred to future work.

\subsubsection{Accreting Black Hole Central Engines}

Perhaps a more appealing scenario for Ic-BL SNe is that they are powered by accretion onto a BH, i.e., the collapsar mechanism \citep{macfadyen:99,kohri:05,kumar:08}. In this model, the inner part of the star where the specific AM $j < j_{\rm ISCO}$ directly collapses to form a Kerr BH. Then, the outer layers with $j>i_{\rm ISCO}$ would circularize and form an accretion disk outside the ISCO. At high accretion rate, the disk may be sufficiently dense and hot to trigger URCA reactions and undergo neutrino cooling \citep{popham:99,narayan:01,dimatteo:02,kohri:05,chen:07}. In the case where the disk is strongly cooled due to neutrino emission, we expect that most of the mass will be viscously accreted to smaller radii \citep{shakura:73}; whereas when the disk is not neutrino-cooled (in the advection-dominated accretion flow, or ADAF regime, \citealt{narayan:94}), then we expect most of the mass to be driven away from the system as disk wind \citep{blandford:99}. The mechanical energy carried by the disk wind can drive a strong stellar explosion, and if this is the case, the supernova emission will be powered by the $^{56}$Ni synthesized in the wind which mixes with the material in the outer envelope of the star during the explosion \citep{macfadyen:99}.


To estimate the energy released through accretion onto the BH, $E_{\rm ac}$, and the energy carried away by a disk wind, $E_{\rm wind}$, we use a simple one-zone disk model as described in Appendix \ref{accretion}. This model evolves the mass/AM of the disk by accounting for material accreting from the envelope of the stellar model, material lost through disk winds, and material falling through the ISCO onto the BH. The middle panel of \autoref{Eengine} shows an estimate of the energy of the disk wind ejecta of our models. In slowly rotating (long orbital period) models, most of the stellar envelope plunges directly into the BH, so only a small amount of mass circularizes outside the ISCO, and the disk wind energy is small. However, many of our rapidly rotating models with short orbital periods $P_{\rm orb} \lesssim 0.6 \, {\rm d}$ have energetic disk winds with energies exceeding $10^{52} \, {\rm erg}$ and are capable of driving Ic-BL SNe. There is little progenitor mass dependence as long as the pre-explosion mass exceeds $\sim \! 6 \, M_\odot$. The amount of mass that circularizes outside $r_{\rm ISCO}$ for models with energetic disk winds is typically $M_{\rm ej} \sim 1-5 \, M_\odot$, most of which is ejected into the wind to form the SN ejecta rather than accreting onto the BH. This ejecta is hydrogen-free and helium-poor, as required to generate a Ic SN. Hence we consider our short-period models to be good candidates for the progenitors of Ic-BL SNe through the collapsar mechanism, but probably not through the millisecond magnetar mechanism. 

Our models may also be good candidates to drive long GRBs via jets generated by the collapsar. To investigate this possibility, we compute the isotropic-equivalent GRB energy $E_{\rm iso}$ of material accreting onto the BH (e.g., \citealt{bavera:21b})
\beq
\label{Eiso}
E_{\rm iso} = \frac{\eta}{f_{\rm B}} E_{\rm ac}
\eeq
where $\eta$ is the fraction of accretion energy channeled into the jet and $f_{\rm B}$ is the beaming fraction. We take $\eta = 0.01$ and $f_{\rm B} = 0.01$ such that $\eta/f_{\rm B} = 1$, but we emphasize there is great uncertainty in these parameters. The top panel of \autoref{Eengine} shows our calculations of $E_{\rm iso}$ for each of our models, along with an approximate detection threshold of $E_{\rm iso} \sim 3 \times 10^{51} \, {\rm erg}$ that lies near the low end of detected GRBs \citep{perley:16}. Our models with $P_{\rm orb} \lesssim 0.75 \, {\rm day}$ and $M_{\rm exp} \gtrsim 5 \, M_\odot$ (i.e., the same models that could generate Ic-BL SNe) are capable of driving long GRBs. More energetic GRBs are predicted to occur from more massive progenitors, with $E_{\rm iso}$ extending up to $\sim \! 10^{54} \, {\rm erg}$ as observed. Since the progenitor models that can produce Ic-BL SNe are the same that produce long GRBs, our models naturally predict the association between long GRBs and Ic-BL SNe.


\subsection{Rates}

We have not performed binary population synthesis so detailed event rate estimates are beyond the scope of this work. Nevertheless, we can crudely extrapolate from observations and existing population synthesis calculations, if we assume that He stars in close binaries are the dominant channel for different types of events. Failure of our models to reach the observed rates may indicate that other channels (e.g., homogeneous evolution of stars spun up by accretion, \citealt{cantiello:07}) are more important. The most recent BH merger rate from LIGO is $R_{\rm BBH} = 23.9^{+14.9}_{-8.6} \, {\rm Gpc}^{-3} \, {\rm yr}^{-1}$ and the NS merger rate is $R_{\rm BNS} = 320^{+490}_{-240} \, {\rm Gpc}^{-3} \, {\rm yr}^{-1}$ \citep{ligopop:21}.

As can be seen from Figure \ref{Pns} and equation \ref{pmax}, the majority of merging NSs likely originate from binaries with pre-explosion orbital period $P_{\rm orb} \lesssim 0.5 \, {\rm d}$ that are efficiently tidally spun-up. Hence, the NS merger rate provides a lower limit on the number of tidally spun-up He stars. Including NS-BH mergers will slightly increase this number. If we assume a Salpeter IMF, then helium stars with $M_{\rm He} \! \gtrsim \! 10 \, M_\odot$ that originate from stars with $M_{\rm ZAMS} \! \gtrsim \! 30 \, M_\odot$ are roughly $30\%$ as common as helium stars with $M_{\rm He} \! \gtrsim \! 4 \, M_\odot$ that originate from $M_{\rm ZAMS} \! \gtrsim \! 12 \, M_\odot$. If these massive He stars ($M_{\rm He} \! \gtrsim \! 10 \, M_\odot$) all form highly magnetized NSs, the event rate of rotating magnetar-powered SNe would be $\sim \! 100 \, {\rm Gpc}^{-3} \, {\rm yr}^{-1}$.

The estimate above can be compared with the observed rate of superluminous Ic SNe of $R_{\rm SLSNe} = 35^{+25}_{-13} \, {\rm Gpc}^{-3} \, {\rm yr}^{-1}$ \citep{frohmaier:21} or $R_{\rm SLSNe} \sim 40 \, {\rm Gpc}^{-3} \, {\rm yr}^{-1}$ \citep{zhao:21}. We thus conclude it is possible for our models to account for the rates of superluminous type Ic SNe. In reality, many of the massive He stars required for rapidly rotating magnetars likely collapse into BHs or produce NSs with too weak/strong fields and would therefore not generate a superluminous Ic SNe. Then again, the rate will be slightly increased by contributions from binaries with $P_{\rm orb} \! \sim \! 0.75 \, {\rm d}$ that are tidally spun up but do not usually merge in a Hubble time. Future work will need to quantify these branching ratios in order to generate better rate estimates.

Our models struggle to match the observed rate of Ic-BL SNe if they are powered by magnetar rotation. While the volumetric rate is uncertain, \cite{shivvers:17b} (see also \citealt{modjaz:20}) estimate that roughly $4\%$ of stripped SNe are type Ic-BL. Given a stripped SNe rate of $R_{\rm Ibc} = 2.4^{+0.8}_{-0.6} \times 10^{4} \, {\rm Gpc}^{-3} \, {\rm yr}^{-1}$ \citep{frohmaier:21}, this entails $R_{\rm Ic-BL} \! \sim \! 10^3 \, {\rm Gpc}^{-3} \, {\rm yr}^{-1}$, roughly 1\% of the total core-collapse SN rate. If only AM in the iron core contributes to the magnetar rotation period, then almost none of our models can reach the $\sim \! 10^{52} \, {\rm erg}$ energy level inferred for type Ic-BL SNe, and we expect $R_{\rm Ic-BL} < 10^2 \, {\rm Gpc}^{-3} \, {\rm yr}^{-1}$, far smaller than observationally inferred. One possibility to reconcile this difference is that fallback accretion very commonly spins up magnetars to millisecond periods, such that some of our less massive and longer orbital period models contribute to the Ic-BL SN population.

If Ic-BL SNe are instead powered by BH accretion via the collapsar mechanism, the same models in \autoref{Eengine} that produce GRBs are those with $E_{\rm ac} \gtrsim 10^{52} \, {\rm erg}$ that can produce Ic-BL SNe, and we might guess the rates of GRBs (after beaming corrections) and Ic-BL SNe to be similar. This is problematic because the uncorrected rate of long GRBs at low redshift is $R_{\rm GRB} \sim 1 \, {\rm Gpc}^{-3} \, {\rm yr}^{-1}$ \citep{wanderman:10}, and using $1/f_B \sim 100$ implies a beaming-corrected volumetric rate of $R_{\rm GRB} \sim \! 100 \, {\rm Gpc}^{-3} \, {\rm yr}^{-1}$. Hence, the observed GRB rate is a factor of $\sim \! 10$ times smaller than that of Ic-BL SNe. It is possible that many GRB jets are choked or that the beaming fraction is smaller than $f_{\rm B} = 0.01$, which could account for the lower observed rate of long GRBs compared to Ic-BL SNe. However, it is difficult for our models to simultaneously reproduce the long GRB rate and the much higher rate of Ic-BL SNe, as explained below.


\cite{bavera:21} performed population synthesis on their stellar models to predict BH merger rates in the range $R_{\rm BBH} \sim \! 25-113 \, {\rm Gpc}^{-3} \, {\rm yr}^{-1}$, approximately consistent with the measured rate from LIGO. \cite{bavera:19} estimate that roughly $40\%$ of these events will have moderate rotation with $\chi_{\rm eff} > 0.1$, which is also similar to the observed fraction of moderately spinning BH mergers \citep{ligoo2b:18}. Since our models predict very similar BH spins as \cite{bavera:19}, we conclude that they can roughly account for the observed rate of of BH mergers, with both low and moderate $\chi_{\rm eff}$. The latter likely have one highly spinning BH like those in \autoref{Eengine} capable of powering a GRB, corresponding to a crude local GRB rate estimate of $R_{\rm GRB} \sim \! 10-40 \, {\rm Gpc}^{-3} \, {\rm yr}^{-1}$ if GRB jets are never choked. This is within a factor of a few of the observed GRB rate above, indicating that this binary channel may be able to account for most long GRBs, the same conclusion reached by \citealt{bavera:21b}.

However, this scenario cannot simultaneously explain the observed rate of Ic-BL SNe, given they are predicted to occur from the same models as those that generate long GRBs and moderate-$\chi_{\rm eff}$ BH mergers. Scaling up the binary collapsar formation rate would help match the Ic-BL SN rate, but it would likely increase the BH merger rate above that observed. One possibility is that Ic-BL SNe are produced by collapsars formed through homogeneous evolution, which may not merge with a companion BH and could therefore increase the Ic-BL rate without increasing the BH merger rate. Another possibility is that most Ic-BL SNe (and perhaps most GRBs) are produced by the first-formed tidally spun-up He star in a short-period ($P \lesssim 1 \, {\rm d}$) binary with a main sequence secondary. The mass transfer at the end of the secondary's life would likely be stable, widening the orbit so that the compact remnant (WD, NS, or BH) would not merge with the first-formed BH within a Hubble time, allowing these systems to enhance the rates of Ic-BL/GRBs without increasing the BH merger rate.

\section{Modeling Uncertainties}
\label{modeling}

Several physical uncertainties affect our compact object spin predictions. Numerical convergence is only a minor source of uncertainty for these models, as shown in \autoref{testing}.

One might expect that the predicted NS and BH spins are sensitive to the initial rotation rate of the progenitor star. However, that is not the case for these models. We ran a set of models in which the main sequence progenitor's rotation rate was boosted to 80\% of its breakup rate at the end of the main sequence (e.g., to simulate accretion from a companion). We then constructed a He star model following the onset of Roche lobe overflow using the same technique described in \autoref{stellar}. However, due to the very strong dependence of the AM viscosity with rotation rate ($\nu_{\rm AM} \propto \Omega^3$), the core rotation rate reaches nearly the same value, regardless of the main sequence star's total AM (see also \citealt{fuller:19,ma:19}). This evidently occurs even for massive stars which evolve very quickly across the HR gap, because we find the He star's initial rotation rate (and subsequent NS and BH spin) is nearly the same, regardless of the spin rate of the main sequence progenitor.

Our models do not include rotational mixing via Eddington-Sweet circulation. To test its importance, we ran a set of models with a generous amount of rotational mixing (\verb|am_D_mix_factor = 1d-1| and \verb|D_ES_factor = 1|) switched on for our helium star models. The resulting NS and BH spin rates are very similar, but slightly smaller for our highest mass models because of slightly increased wind mass loss rates in these models. 

A major assumption of our predictions for NS rotation rates is that of AM conservation during the core-collapse explosion. Multi-dimensional simulations \citep{muller:18,chan:20,stockinger:20,janka:21} instead show that that asymmetric explosions and fallback accretion can significantly alter the spin rate of the NS, spinning it up to spin periods as short as milliseconds. However, those simulations may also overestimate the spin-up via fallback because of numerical gridding effects or if fallback material is inefficiently accreted \citep{janka:21}. Given the long NS spin periods predicted for our low-mass progenitor models, it is likely that the true spin periods for those NSs are set by fallback and are shorter than our predictions. However, given the rarity of engine-powered SNe, we are skeptical that a common process like asymmetric fallback accretion can explain the millisecond periods needed to power engine-driven transients, nor why those transients are associated with hydrogen-free progenitor stars.

\subsection{AM transport Uncertainties}

Despite asteroseismically measured rotation rates of low-mass red giants, the physical mechanisms behind AM transport are not well understood, translating to substantial uncertainties for predictions of core rotation rates in high-mass stars. Our updated AM transport prescription based on the Tayler instability described in Section \ref{angmom} yields similar predictions for low-mass red giants as the original prescription of \cite{fuller:19}, and it is therefore asteroseimically calibrated. However, this underpredicts the core rotation rates of low-mass sub-giant stars \citep{eggenberger:19}, and it overpredicts the core rotation rates of secondary clump stars \citep{tayar:19}. While our models match the data much better than models including the TS dynamo \citep{cantiello:14}, they are clearly not perfect. 

Moreover, the observed scatter \citep{gehan:18} in core rotation rates by a factor of $\sim \!3$ above and below the predictions of \cite{fuller:19} for low-mass red giants likely entails diversity in the effective AM transport efficiency factor $\alpha$. A similar scatter might be expected for high-mass stars. We find that the predicted NS rotation period scales roughly linearly with $\alpha$ for the models of this paper, with larger values of $\alpha$ creating more efficient AM transport and slower NS spin. Hence, real stars may exhibit a scatter in NS rotation periods by a factor of a few above or below our predictions. This may allow slightly lower mass stars to contribute substantially to rotation-powered energetic transients.

Our updated prescription predicts NS rotation periods roughly half that of \cite{fuller:19} when applied to the H-poor stellar models of this paper. The difference is smaller for high-mass models but is slightly larger for our lowest mass models. For comparison, we ran several of our models with the original TS dynamo prescription of \citep{spruit:02}. This predicts NS rotation rates $\sim \!4$ times faster than our low-mass models, with $P_{\rm NS} \! \sim \! 10 \, {\rm ms}$ for $M_{\rm exp} \lesssim 5 \, M_\odot$. Interestingly, however, both models predict similar NS rotation rates from high-mass stars. Hence, multiple proposed AM transport prescriptions coincidentally predict $P_{\rm NS} \! \sim \! 1 \, {\rm ms}$ for short-period binaries with $M_{\rm exp} \gtrsim 15 \, M_\odot$. For BH spin, both this work and models employing the TS dynamo now predict typical spins of $a_{\rm BH} \sim 0.03$ from BHs born from He stars in wide binaries \citep{qin:18,bavera:19}. This statement is dependent on the progenitor mass, rotation rate, metallicity, and mass loss prescription, so it is largely coincidental.

In contrast to our models, \cite{kissin:15} proposed that magnetic torques enforce nearly rigid rotation in radiative regions, while convective AM pumping causes differential rotation in thick convective layers (see also \citealt{takahashi:21}).
\cite{kissin:18} examined the core AM of massive stars in this scenario, predicting slowly rotating NSs in single stars, but finding that tidal spin-up of a red supergiant progenitor can lead to a rapidly rotating NS or BH. In their model, rapid rotation is possible for hydrogen-rich progenitors, in contrast to our models. They did not examine tidal spin-up of helium stars, nor did they investigate mass/metallicity dependence in detail. Future work should examine the convective AM pumping scenario to make distinguishing predictions from other models. 


\subsection{Helium stars from other evolutionary channels}
\label{otherhe}

\subsubsection{Stable mass transfer}
\label{stable}

Several recent works (e.g., \citealt{pavlovskii:17,klencki:20,vanson:21,marchant:21})
have argued that many merging BHs may form via orbital decay during stable mass transfer rather than common envelope evolution. Our models are agnostic to the progenitor channel as long as it produces a nearly hydrogen-free progenitor star in a short-period orbit, and as long as the envelope stripping occurs on a very short time scale. The stable mass transfer scenario may leave a thicker hydrogen shell than used in our models, and it may proceed over a slightly longer (thermal) time scale than the envelope stripping prescription that we use, so the resultant NS and BH spins may be somewhat different.

To explore the potential effect of a thin hydrogen envelope, we ran a few models identical to those above, but leaving a couple tenths of a solar mass of hydrogen after the stripping process. The results for high-mass stars ($M_{\rm ZAMS} \gtrsim 40 \, M_\odot$) are nearly unaffected because the residual hydrogen is quickly lost by winds. For lower mass stars, however, the hydrogen envelopes can be retained through helium core burning, causing the star to greatly inflate during helium shell burning (e.g., \citealt{laplace:20}). The expanding hydrogen envelope extracts AM from the core, most of which is lost because this hydrogen envelope overflows the Roche lobe and is lost via case BB mass transfer (e.g., \citealt{tauris:15}). In our models, this can cause the BH spin to be more than 50\% smaller, and in certain cases could also cause the NS spin period to be longer. Future work should investigate this scenario more thoroughly.

\subsubsection{Helium stars from chemically homogeneous evolution}
\label{homogeneous}

We have not examined massive helium stars formed through the homogeneous evolution channel \citep{maeder:87,woosley:06,yoon:06,cantiello:07}, in which there is no hydrogen envelope to absorb the core's AM. \cite{aguileradena:18} and \cite{aguileradena:20} examined the evolution of low-metallicity helium stars evolving through homogeneous evolution, adopting the TS dynamo \citep{spruit:02} for AM transport. Those models also implemented rotational mixing enhanced by a factor of 10, such that they burned nearly all their helium and evolved into nearly homogeneous carbon-oxygen stars. Because those models did not include tidal spin-up, their total AM is determined by their initial spin rate and mass loss history. 

The AM of the iron core, however, is determined primarily by AM exchange with the overlying carbon/oxygen core and thus depends on the AM transport prescription. Our most massive tidally spun-up models ($M_{\rm exp} \! \sim \! 20 \, M_\odot$) have average core specific AM (within the inner $1.6 \, M_\odot$) of $j\approx 3 \times 10^{15} \, {\rm cm}^2 \, {\rm s}^{-1}$, similar to the models of \cite{aguileradena:20}. However, our low-mass models ($M_{\rm exp} \lesssim 5 M_\odot$) only have specific AM of $j \! \sim \! 10^{14} \, {\rm cm}^2 \, {\rm s}^{-1}$, several times smaller than those of \cite{aguileradena:20}. There are two reasons for this. First, the enhanced mixing of \cite{aguileradena:20} produces larger carbon-oxygen cores and more compact stars without extended helium envelopes, reducing the amount of AM transferred from the core to the helium envelope. Second, the less efficient AM transport from the TS dynamo extracts less AM from the core. We believe the core rotation rates of \cite{aguileradena:20} (especially for their low-mass models) could be overestimated because the TS dynamo under-predicts AM transport efficiency and over-predicts the core spin rates of low-mass red giants \citep{cantiello:14,fuller:19}. Additionally, the enhanced rotational mixing does not appear to be supported by measurements of nitrogen abundances of rotating stars \citep{brott:11}.

\section{Conclusion}

We have generated a suite of detailed binary stellar evolution models, examining angular momentum (AM) transport and core rotation rates of massive helium stars in close binary systems. Such binaries could possibly be formed via common envelope evolution (e.g., \citealt{belczynski:02}), or via stable mass transfer (e.g., \citealt{marchant:21}), which may be more likely for massive helium stars \citep{klencki:20}. Our models improve upon prior efforts by implementing an updated AM transport prescription based on magnetic torques associated with the Tayler instability \citep{fuller:19}, which has been calibrated with asteroseismic core rotation rates of low-mass red giant stars. The models also include tidal torques from the companion star which greatly increase the AM content for binaries with orbital periods $P_{\rm orb} \lesssim 1 \, {\rm d}$, allowing for rapid core rotation that could power various types of SNe and energetic transients. The efficient AM transport of our models allows most of the iron core's AM to be extracted before core-collapse, predicting slower core rotation than most prior models.

Based on the core AM content just before core-collapse, we predict the rotation rates of neutron stars (NSs) born from binary helium stars (\autoref{Pns}), assuming AM is conserved during collapse. We find that NSs born from low-mass progenitors ($M_{\rm exp} \lesssim 5 \, M_\odot$) and wide binaries $(P_{\rm orb} \gtrsim 1 \, {\rm d}$) are generally slowly rotating, with initial rotation rates $P_{\rm NS} \gtrsim 10 \, {\rm ms}$. The rotation periods of these NSs may instead be determined by asymmetric fallback accretion (e.g., \citealt{chan:20,stockinger:20}). NSs born from massive progenitors in close binaries can be rapidly rotating, with $P_{\rm NS} \! \sim \! 1 \, {\rm ms}$ for the most massive progenitors in very tight binaries. Core contraction of very massive stars occurs so quickly that there is not enough time to remove the core AM before core-collapse, so our models predict a strong correlation between the NS rotation rate and progenitor/ejecta mass, with approximate scaling $P_{\rm NS} \propto M_{\rm exp}^{-1.4} P_{\rm orb}^{0.8}$ for stars in tight binaries. A similar correlation has been observationally inferred from magnetar models for superluminous type Ic SNe \citep{blanchard:20}, though our predicted ejecta masses are several times larger than many current estimates for observed events.

When a black hole (BH) is formed upon core-collapse, our models predict low spin $(a \lesssim 0.1$) for BHs in wide binaries with $P_{\rm orb} \gtrsim 1 \, {\rm d}$ (\autoref{abh}). However, tidal spin-up of the progenitor at shorter periods allows for rapid rotation, with $a \! \sim \! 1$ possible at very short orbital periods, and an approximate scaling $a \propto M_{\rm BH}^{0.5} P_{\rm orb}^{-2.5}$. Unlike NSs, the BH spin is nearly independent of AM transport and our models yield results similar to other recent work (e.g., \citealt{qin:18,bavera:19}). These models predict low values of $\chi_{\rm eff}$ for most BH mergers observed by LIGO, but with a significant fraction of moderate $\chi_{\rm eff}$ events arising from BHs in close binaries. This appears consistent with the low/moderate measured values of $\chi_{\rm eff}$ for most LIGO events thus far.

Finally, we investigate the possibility of generating energetic transients (superluminous type Ic SNe, type Ic-BL SNe, and GRBs) via rotational or accretion power in our models. The AM content of the envelopes of our short-period binaries are likely sufficient to power GRBs via the collapsar model (\autoref{Eengine}). Using only the AM in the iron core, our massive models can create rapidly rotating magnetars capable of powering superluminous SNe, but only for very massive stars in very tight binaries with an appropriately tuned magnetar field strength, so it is unclear if they can account for the observed rates. This is more problematic for magnetar powering of Ic-BL SNe, which are more common and have larger explosion energies of $\sim \! 10^{52} \, {\rm erg}$. These events require significant fallback accretion and spin-up of the magnetar to achieve the required rotational energy, or they must be powered by BH accretion rather than magnetar spin-down. While our models capable of powering GRBs are also predicted to power $\sim \! 10^{52} \, {\rm erg}$ SNe Ic-BL via outflows from accretion disk winds, it is difficult to account for the much higher observed rate of SNe Ic-BL.

\section*{Acknowledgments}

JF is thankful for support through an Innovator Grant from The Rose Hills Foundation, and the Sloan Foundation through grant FG-2018-10515. WL is supported by the Lyman Spitzer, Jr. Fellowship at Princeton University.

\section*{Data Availability}
Data and source code is available upon request to the authors.

\bibliography{CoreRotBib,library}

\begin{thebibliography}{}
\makeatletter
\relax
\def\mn@urlcharsother{\let\do\@makeother \do\$\do\&\do\#\do\^\do\_\do\%\do\~}
\def\mn@doi{\begingroup\mn@urlcharsother \@ifnextchar [ {\mn@doi@}
  {\mn@doi@[]}}
\def\mn@doi@[#1]#2{\def\@tempa{#1}\ifx\@tempa\@empty \href
  {http://dx.doi.org/#2} {doi:#2}\else \href {http://dx.doi.org/#2} {#1}\fi
  \endgroup}
\def\mn@eprint#1#2{\mn@eprint@#1:#2::\@nil}
\def\mn@eprint@arXiv#1{\href {http://arxiv.org/abs/#1} {{\tt arXiv:#1}}}
\def\mn@eprint@dblp#1{\href {http://dblp.uni-trier.de/rec/bibtex/#1.xml}
  {dblp:#1}}
\def\mn@eprint@#1:#2:#3:#4\@nil{\def\@tempa {#1}\def\@tempb {#2}\def\@tempc
  {#3}\ifx \@tempc \@empty \let \@tempc \@tempb \let \@tempb \@tempa \fi \ifx
  \@tempb \@empty \def\@tempb {arXiv}\fi \@ifundefined
  {mn@eprint@\@tempb}{\@tempb:\@tempc}{\expandafter \expandafter \csname
  mn@eprint@\@tempb\endcsname \expandafter{\@tempc}}}

\bibitem[\protect\citeauthoryear{{Abbott} et~al.,}{{Abbott}
  et~al.}{2021}]{ligopop:21}
{Abbott} R.,  et~al., 2021, \mn@doi [\apjl] {10.3847/2041-8213/abe949}, \href
  {https://ui.adsabs.harvard.edu/abs/2021ApJ...913L...7A} {913, L7}

\bibitem[\protect\citeauthoryear{{Afsariardchi}, {Drout}, {Khatami}, {Matzner},
  {Moon}  \& {Ni}}{{Afsariardchi} et~al.}{2020}]{afsariardchi:21}
{Afsariardchi} N.,  {Drout} M.~R.,  {Khatami} D.,  {Matzner} C.~D.,  {Moon}
  D.-S.,   {Ni} Y.~Q.,  2020, arXiv e-prints, \href
  {https://ui.adsabs.harvard.edu/abs/2020arXiv200906683A} {p. arXiv:2009.06683}

\bibitem[\protect\citeauthoryear{{Aguilera-Dena}, {Langer}, {Moriya}  \&
  {Schootemeijer}}{{Aguilera-Dena} et~al.}{2018}]{aguileradena:18}
{Aguilera-Dena} D.~R.,  {Langer} N.,  {Moriya} T.~J.,   {Schootemeijer} A.,
  2018, \mn@doi [\apj] {10.3847/1538-4357/aabfc1}, \href
  {http://adsabs.harvard.edu/abs/2018ApJ...858..115A} {858, 115}

\bibitem[\protect\citeauthoryear{{Aguilera-Dena}, {Langer}, {Antoniadis}  \&
  {M{\"u}ller}}{{Aguilera-Dena} et~al.}{2020}]{aguileradena:20}
{Aguilera-Dena} D.~R.,  {Langer} N.,  {Antoniadis} J.,   {M{\"u}ller} B.,
  2020, \mn@doi [\apj] {10.3847/1538-4357/abb138}, \href
  {https://ui.adsabs.harvard.edu/abs/2020ApJ...901..114A} {901, 114}

\bibitem[\protect\citeauthoryear{{Aloy} \& {Obergaulinger}}{{Aloy} \&
  {Obergaulinger}}{2021}]{aloy:21}
{Aloy} M.~{\'A}.,  {Obergaulinger} M.,  2021, \mn@doi [\mnras]
  {10.1093/mnras/staa3273}, \href
  {https://ui.adsabs.harvard.edu/abs/2021MNRAS.500.4365A} {500, 4365}

\bibitem[\protect\citeauthoryear{{Arnett}}{{Arnett}}{1980}]{arnett:80}
{Arnett} W.~D.,  1980, \mn@doi [\apj] {10.1086/157898}, \href
  {http://adsabs.harvard.edu/abs/1980ApJ...237..541A} {237, 541}

\bibitem[\protect\citeauthoryear{{Barnes}, {Duffell}, {Liu}, {Modjaz},
  {Bianco}, {Kasen}  \& {MacFadyen}}{{Barnes} et~al.}{2018}]{barnes:18}
{Barnes} J.,  {Duffell} P.~C.,  {Liu} Y.,  {Modjaz} M.,  {Bianco} F.~B.,
  {Kasen} D.,   {MacFadyen} A.~I.,  2018, \mn@doi [\apj]
  {10.3847/1538-4357/aabf84}, \href
  {https://ui.adsabs.harvard.edu/abs/2018ApJ...860...38B} {860, 38}

\bibitem[\protect\citeauthoryear{{Batta} \& {Ramirez-Ruiz}}{{Batta} \&
  {Ramirez-Ruiz}}{2019}]{batta:19}
{Batta} A.,  {Ramirez-Ruiz} E.,  2019, arXiv e-prints, \href
  {https://ui.adsabs.harvard.edu/abs/2019arXiv190404835B} {p. arXiv:1904.04835}

\bibitem[\protect\citeauthoryear{{Batta}, {Ramirez-Ruiz}  \& {Fryer}}{{Batta}
  et~al.}{2017}]{batta:17}
{Batta} A.,  {Ramirez-Ruiz} E.,   {Fryer} C.,  2017, \mn@doi [\apjl]
  {10.3847/2041-8213/aa8506}, \href
  {https://ui.adsabs.harvard.edu/abs/2017ApJ...846L..15B} {846, L15}

\bibitem[\protect\citeauthoryear{{Bavera} et~al.,}{{Bavera}
  et~al.}{2019}]{bavera:19}
{Bavera} S.~S.,  et~al., 2019, arXiv e-prints, \href
  {https://ui.adsabs.harvard.edu/abs/2019arXiv190612257B} {p. arXiv:1906.12257}

\bibitem[\protect\citeauthoryear{{Bavera} et~al.,}{{Bavera}
  et~al.}{2021a}]{bavera:21b}
{Bavera} S.~S.,  et~al., 2021a, arXiv e-prints, \href
  {https://ui.adsabs.harvard.edu/abs/2021arXiv210615841B} {p. arXiv:2106.15841}

\bibitem[\protect\citeauthoryear{{Bavera} et~al.,}{{Bavera}
  et~al.}{2021b}]{bavera:21}
{Bavera} S.~S.,  et~al., 2021b, \mn@doi [\aap] {10.1051/0004-6361/202039804},
  \href {https://ui.adsabs.harvard.edu/abs/2021A&A...647A.153B} {647, A153}

\bibitem[\protect\citeauthoryear{{Beck} et~al.,}{{Beck} et~al.}{2012}]{beck:12}
{Beck} P.~G.,  et~al., 2012, \mn@doi [\nat] {10.1038/nature10612}, \href
  {http://adsabs.harvard.edu/abs/2012Natur.481...55B} {481, 55}

\bibitem[\protect\citeauthoryear{{Belczynski, K.} et~al.,}{{Belczynski, K.}
  et~al.}{2020}]{belczynski:20}
{Belczynski, K.} et~al., 2020, \mn@doi [A\&A] {10.1051/0004-6361/201936528},
  636, A104

\bibitem[\protect\citeauthoryear{{Belczynski}, {Kalogera}  \&
  {Bulik}}{{Belczynski} et~al.}{2002}]{belczynski:02}
{Belczynski} K.,  {Kalogera} V.,   {Bulik} T.,  2002, \mn@doi [\apj]
  {10.1086/340304}, \href
  {https://ui.adsabs.harvard.edu/abs/2002ApJ...572..407B} {572, 407}

\bibitem[\protect\citeauthoryear{{Belczynski}, {Done}  \&
  {Lasota}}{{Belczynski} et~al.}{2021}]{belczynski:21}
{Belczynski} K.,  {Done} C.,   {Lasota} J.~P.,  2021, arXiv e-prints, \href
  {https://ui.adsabs.harvard.edu/abs/2021arXiv211109401B} {p. arXiv:2111.09401}

\bibitem[\protect\citeauthoryear{{Belkacem} et~al.,}{{Belkacem}
  et~al.}{2015}]{belkacem:15}
{Belkacem} K.,  et~al., 2015, \mn@doi [\aap] {10.1051/0004-6361/201526043},
  \href {http://adsabs.harvard.edu/abs/2015A%26A...579A..31B} {579, A31}

\bibitem[\protect\citeauthoryear{{Beniamini}, {Giannios}  \&
  {Metzger}}{{Beniamini} et~al.}{2017}]{beniamini:17}
{Beniamini} P.,  {Giannios} D.,   {Metzger} B.~D.,  2017, \mn@doi [\mnras]
  {10.1093/mnras/stx2095}, \href
  {https://ui.adsabs.harvard.edu/abs/2017MNRAS.472.3058B} {472, 3058}

\bibitem[\protect\citeauthoryear{{Blaauw}}{{Blaauw}}{1961}]{blaauw:61}
{Blaauw} A.,  1961, \bain, \href
  {https://ui.adsabs.harvard.edu/abs/1961BAN....15..265B} {15, 265}

\bibitem[\protect\citeauthoryear{{Blanchard}, {Berger}, {Nicholl}  \&
  {Villar}}{{Blanchard} et~al.}{2020}]{blanchard:20}
{Blanchard} P.~K.,  {Berger} E.,  {Nicholl} M.,   {Villar} V.~A.,  2020,
  \mn@doi [\apj] {10.3847/1538-4357/ab9638}, \href
  {https://ui.adsabs.harvard.edu/abs/2020ApJ...897..114B} {897, 114}

\bibitem[\protect\citeauthoryear{{Blandford} \& {Begelman}}{{Blandford} \&
  {Begelman}}{1999}]{blandford:99}
{Blandford} R.~D.,  {Begelman} M.~C.,  1999, \mn@doi [\mnras]
  {10.1046/j.1365-8711.1999.02358.x}, \href
  {https://ui.adsabs.harvard.edu/abs/1999MNRAS.303L...1B} {303, L1}

\bibitem[\protect\citeauthoryear{{Brott} et~al.,}{{Brott}
  et~al.}{2011}]{brott:11}
{Brott} I.,  et~al., 2011, \mn@doi [\aap] {10.1051/0004-6361/201016114}, \href
  {https://ui.adsabs.harvard.edu/abs/2011A&A...530A.116B} {530, A116}

\bibitem[\protect\citeauthoryear{{Cantiello}, {Yoon}, {Langer}  \&
  {Livio}}{{Cantiello} et~al.}{2007}]{cantiello:07}
{Cantiello} M.,  {Yoon} S.~C.,  {Langer} N.,   {Livio} M.,  2007, \mn@doi
  [\aap] {10.1051/0004-6361:20077115}, \href
  {https://ui.adsabs.harvard.edu/abs/2007A&A...465L..29C} {465, L29}

\bibitem[\protect\citeauthoryear{{Cantiello}, {Mankovich}, {Bildsten},
  {Christensen-Dalsgaard}  \& {Paxton}}{{Cantiello}
  et~al.}{2014}]{cantiello:14}
{Cantiello} M.,  {Mankovich} C.,  {Bildsten} L.,  {Christensen-Dalsgaard} J.,
  {Paxton} B.,  2014, \mn@doi [\apj] {10.1088/0004-637X/788/1/93}, \href
  {http://adsabs.harvard.edu/abs/2014ApJ...788...93C} {788, 93}

\bibitem[\protect\citeauthoryear{{Chan}, {M{\"u}ller}  \& {Heger}}{{Chan}
  et~al.}{2020}]{chan:20}
{Chan} C.,  {M{\"u}ller} B.,   {Heger} A.,  2020, \mn@doi [\mnras]
  {10.1093/mnras/staa1431}, \href
  {https://ui.adsabs.harvard.edu/abs/2020MNRAS.495.3751C} {495, 3751}

\bibitem[\protect\citeauthoryear{{Chen} \& {Beloborodov}}{{Chen} \&
  {Beloborodov}}{2007}]{chen:07}
{Chen} W.-X.,  {Beloborodov} A.~M.,  2007, \mn@doi [\apj] {10.1086/508923},
  \href {https://ui.adsabs.harvard.edu/abs/2007ApJ...657..383C} {657, 383}

\bibitem[\protect\citeauthoryear{{Deheuvels} et~al.,}{{Deheuvels}
  et~al.}{2014}]{deheuvels:14}
{Deheuvels} S.,  et~al., 2014, \mn@doi [\aap] {10.1051/0004-6361/201322779},
  \href {http://adsabs.harvard.edu/abs/2014A%26A...564A..27D} {564, A27}

\bibitem[\protect\citeauthoryear{{Deheuvels}, {Ballot}, {Beck}, {Mosser},
  {{\O}stensen}, {Garc{\'{\i}}a}  \& {Goupil}}{{Deheuvels}
  et~al.}{2015}]{deheuvels:15}
{Deheuvels} S.,  {Ballot} J.,  {Beck} P.~G.,  {Mosser} B.,  {{\O}stensen} R.,
  {Garc{\'{\i}}a} R.~A.,   {Goupil} M.~J.,  2015, \mn@doi [\aap]
  {10.1051/0004-6361/201526449}, \href
  {http://adsabs.harvard.edu/abs/2015A%26A...580A..96D} {580, A96}

\bibitem[\protect\citeauthoryear{{Di Matteo}, {Perna}  \& {Narayan}}{{Di
  Matteo} et~al.}{2002}]{dimatteo:02}
{Di Matteo} T.,  {Perna} R.,   {Narayan} R.,  2002, \mn@doi [\apj]
  {10.1086/342832}, \href
  {https://ui.adsabs.harvard.edu/abs/2002ApJ...579..706D} {579, 706}

\bibitem[\protect\citeauthoryear{{Eggenberger} et~al.,}{{Eggenberger}
  et~al.}{2017}]{eggenberger:17}
{Eggenberger} P.,  et~al., 2017, \mn@doi [\aap] {10.1051/0004-6361/201629459},
  \href {http://adsabs.harvard.edu/abs/2017A%26A...599A..18E} {599, A18}

\bibitem[\protect\citeauthoryear{{Eggenberger}, {den Hartogh}, {Buldgen},
  {Meynet}, {Salmon}  \& {Deheuvels}}{{Eggenberger}
  et~al.}{2019}]{eggenberger:19}
{Eggenberger} P.,  {den Hartogh} J.~W.,  {Buldgen} G.,  {Meynet} G.,  {Salmon}
  S.~J.~A.~J.,   {Deheuvels} S.,  2019, \mn@doi [\aap]
  {10.1051/0004-6361/201936348}, \href
  {https://ui.adsabs.harvard.edu/abs/2019A&A...631L...6E} {631, L6}

\bibitem[\protect\citeauthoryear{{Ertl}, {Woosley}, {Sukhbold}  \&
  {Janka}}{{Ertl} et~al.}{2020}]{ertl:20}
{Ertl} T.,  {Woosley} S.~E.,  {Sukhbold} T.,   {Janka} H.~T.,  2020, \mn@doi
  [\apj] {10.3847/1538-4357/ab6458}, \href
  {https://ui.adsabs.harvard.edu/abs/2020ApJ...890...51E} {890, 51}

\bibitem[\protect\citeauthoryear{{Faucher-Gigu{\`e}re} \&
  {Kaspi}}{{Faucher-Gigu{\`e}re} \& {Kaspi}}{2006}]{faucher:06}
{Faucher-Gigu{\`e}re} C.-A.,  {Kaspi} V.~M.,  2006, \mn@doi [\apj]
  {10.1086/501516}, \href {http://adsabs.harvard.edu/abs/2006ApJ...643..332F}
  {643, 332}

\bibitem[\protect\citeauthoryear{{Fern{\'a}ndez}, {Quataert}, {Kashiyama}  \&
  {Coughlin}}{{Fern{\'a}ndez} et~al.}{2018}]{fernandez:18}
{Fern{\'a}ndez} R.,  {Quataert} E.,  {Kashiyama} K.,   {Coughlin} E.~R.,  2018,
  \mn@doi [\mnras] {10.1093/mnras/sty306}, \href
  {http://adsabs.harvard.edu/abs/2018MNRAS.476.2366F} {476, 2366}

\bibitem[\protect\citeauthoryear{{Fishbach} \& {Kalogera}}{{Fishbach} \&
  {Kalogera}}{2021}]{fishbach:21}
{Fishbach} M.,  {Kalogera} V.,  2021, arXiv e-prints, \href
  {https://ui.adsabs.harvard.edu/abs/2021arXiv211102935F} {p. arXiv:2111.02935}

\bibitem[\protect\citeauthoryear{{Frohmaier} et~al.,}{{Frohmaier}
  et~al.}{2021}]{frohmaier:21}
{Frohmaier} C.,  et~al., 2021, \mn@doi [\mnras] {10.1093/mnras/staa3607}, \href
  {https://ui.adsabs.harvard.edu/abs/2021MNRAS.500.5142F} {500, 5142}

\bibitem[\protect\citeauthoryear{{Fryer}, {Lloyd-Ronning}, {Wollaeger},
  {Wiggins}, {Miller}, {Dolence}, {Ryan}  \& {Fields}}{{Fryer}
  et~al.}{2019}]{fryer:19}
{Fryer} C.~L.,  {Lloyd-Ronning} N.,  {Wollaeger} R.,  {Wiggins} B.,  {Miller}
  J.,  {Dolence} J.,  {Ryan} B.,   {Fields} C.~E.,  2019, arXiv e-prints, \href
  {http://adsabs.harvard.edu/abs/2019arXiv190410008F} {}

\bibitem[\protect\citeauthoryear{{Fuller} \& {Ma}}{{Fuller} \&
  {Ma}}{2019}]{fullerma:19}
{Fuller} J.,  {Ma} L.,  2019, \mn@doi [\apjl] {10.3847/2041-8213/ab339b}, \href
  {https://ui.adsabs.harvard.edu/abs/2019ApJ...881L...1F} {881, L1}

\bibitem[\protect\citeauthoryear{{Fuller}, {Lecoanet}, {Cantiello}  \&
  {Brown}}{{Fuller} et~al.}{2014}]{fullerwave:14}
{Fuller} J.,  {Lecoanet} D.,  {Cantiello} M.,   {Brown} B.,  2014, \mn@doi
  [\apj] {10.1088/0004-637X/796/1/17}, \href
  {http://adsabs.harvard.edu/abs/2014ApJ...796...17F} {796, 17}

\bibitem[\protect\citeauthoryear{{Fuller}, {Cantiello}, {Lecoanet}  \&
  {Quataert}}{{Fuller} et~al.}{2015}]{fullerwave:15}
{Fuller} J.,  {Cantiello} M.,  {Lecoanet} D.,   {Quataert} E.,  2015, \mn@doi
  [\apj] {10.1088/0004-637X/810/2/101}, \href
  {http://adsabs.harvard.edu/abs/2015ApJ...810..101F} {810, 101}

\bibitem[\protect\citeauthoryear{{Fuller}, {Piro}  \& {Jermyn}}{{Fuller}
  et~al.}{2019}]{fuller:19}
{Fuller} J.,  {Piro} A.~L.,   {Jermyn} A.~S.,  2019, \mn@doi [\mnras]
  {10.1093/mnras/stz514}, \href
  {https://ui.adsabs.harvard.edu/abs/2019MNRAS.485.3661F} {485, 3661}

\bibitem[\protect\citeauthoryear{{Gehan}, {Mosser}, {Michel}, {Samadi}  \&
  {Kallinger}}{{Gehan} et~al.}{2018}]{gehan:18}
{Gehan} C.,  {Mosser} B.,  {Michel} E.,  {Samadi} R.,   {Kallinger} T.,  2018,
  \mn@doi [\aap] {10.1051/0004-6361/201832822}, \href
  {http://adsabs.harvard.edu/abs/2018A%26A...616A..24G} {616, A24}

\bibitem[\protect\citeauthoryear{{Georgy}, {Ekstr{\"o}m}, {Granada}, {Meynet},
  {Mowlavi}, {Eggenberger}  \& {Maeder}}{{Georgy} et~al.}{2013}]{georgy:13}
{Georgy} C.,  {Ekstr{\"o}m} S.,  {Granada} A.,  {Meynet} G.,  {Mowlavi} N.,
  {Eggenberger} P.,   {Maeder} A.,  2013, \mn@doi [\aap]
  {10.1051/0004-6361/201220558}, \href
  {http://adsabs.harvard.edu/abs/2013A%26A...553A..24G} {553, A24}

\bibitem[\protect\citeauthoryear{{Gull{\'o}n}, {Miralles}, {Vigan{\`o}}  \&
  {Pons}}{{Gull{\'o}n} et~al.}{2014}]{gullon:14}
{Gull{\'o}n} M.,  {Miralles} J.~A.,  {Vigan{\`o}} D.,   {Pons} J.~A.,  2014,
  \mn@doi [\mnras] {10.1093/mnras/stu1253}, \href
  {http://adsabs.harvard.edu/abs/2014MNRAS.443.1891G} {443, 1891}

\bibitem[\protect\citeauthoryear{{Guti{\'e}rrez} et~al.,}{{Guti{\'e}rrez}
  et~al.}{2021}]{gutierrez:21}
{Guti{\'e}rrez} C.~P.,  et~al., 2021, \mn@doi [\mnras]
  {10.1093/mnras/stab1009}, \href
  {https://ui.adsabs.harvard.edu/abs/2021MNRAS.504.4907G} {504, 4907}

\bibitem[\protect\citeauthoryear{{Heger}, {Langer}  \& {Woosley}}{{Heger}
  et~al.}{2000}]{heger:00}
{Heger} A.,  {Langer} N.,   {Woosley} S.~E.,  2000, \mn@doi [\apj]
  {10.1086/308158}, \href {http://adsabs.harvard.edu/abs/2000ApJ...528..368H}
  {528, 368}

\bibitem[\protect\citeauthoryear{{Heger}, {Woosley}  \& {Spruit}}{{Heger}
  et~al.}{2005}]{heger:05}
{Heger} A.,  {Woosley} S.~E.,   {Spruit} H.~C.,  2005, \mn@doi [\apj]
  {10.1086/429868}, \href {http://adsabs.harvard.edu/abs/2005ApJ...626..350H}
  {626, 350}

\bibitem[\protect\citeauthoryear{{Hermes} et~al.,}{{Hermes}
  et~al.}{2017}]{hermes:17}
{Hermes} J.~J.,  et~al., 2017, \mn@doi [\apjs] {10.3847/1538-4365/aa8bb5},
  \href {http://adsabs.harvard.edu/abs/2017ApJS..232...23H} {232, 23}

\bibitem[\protect\citeauthoryear{{Janka}, {Wongwathanarat}  \&
  {Kramer}}{{Janka} et~al.}{2021}]{janka:21}
{Janka} H.~T.,  {Wongwathanarat} A.,   {Kramer} M.,  2021, arXiv e-prints,
  \href {https://ui.adsabs.harvard.edu/abs/2021arXiv210407493J} {p.
  arXiv:2104.07493}

\bibitem[\protect\citeauthoryear{Kasen \& Bildsten}{Kasen \&
  Bildsten}{2010}]{kasen:10}
Kasen D.,  Bildsten L.,  2010, \mn@doi [{ApJ}] {10.1088/0004-637x/717/1/245},
  717, 245

\bibitem[\protect\citeauthoryear{{Khatami} \& {Kasen}}{{Khatami} \&
  {Kasen}}{2019}]{khatami:19}
{Khatami} D.~K.,  {Kasen} D.~N.,  2019, \mn@doi [\apj]
  {10.3847/1538-4357/ab1f09}, \href
  {https://ui.adsabs.harvard.edu/abs/2019ApJ...878...56K} {878, 56}

\bibitem[\protect\citeauthoryear{{Kissin} \& {Thompson}}{{Kissin} \&
  {Thompson}}{2015}]{kissin:15}
{Kissin} Y.,  {Thompson} C.,  2015, \mn@doi [\apj]
  {10.1088/0004-637X/808/1/35}, \href
  {http://adsabs.harvard.edu/abs/2015ApJ...808...35K} {808, 35}

\bibitem[\protect\citeauthoryear{{Kissin} \& {Thompson}}{{Kissin} \&
  {Thompson}}{2018}]{kissin:18}
{Kissin} Y.,  {Thompson} C.,  2018, \mn@doi [\apj] {10.3847/1538-4357/aab1fb},
  \href {http://adsabs.harvard.edu/abs/2018ApJ...862..111K} {862, 111}

\bibitem[\protect\citeauthoryear{{Klencki}, {Nelemans}, {Istrate}  \&
  {Pols}}{{Klencki} et~al.}{2020}]{klencki:20}
{Klencki} J.,  {Nelemans} G.,  {Istrate} A.~G.,   {Pols} O.,  2020, \mn@doi
  [\aap] {10.1051/0004-6361/202037694}, \href
  {https://ui.adsabs.harvard.edu/abs/2020A&A...638A..55K} {638, A55}

\bibitem[\protect\citeauthoryear{{Kohri}, {Narayan}  \& {Piran}}{{Kohri}
  et~al.}{2005}]{kohri:05}
{Kohri} K.,  {Narayan} R.,   {Piran} T.,  2005, \mn@doi [\apj]
  {10.1086/431354}, \href
  {https://ui.adsabs.harvard.edu/abs/2005ApJ...629..341K} {629, 341}

\bibitem[\protect\citeauthoryear{{K{\"o}nyves-T{\'o}th} \&
  {Vink{\'o}}}{{K{\"o}nyves-T{\'o}th} \& {Vink{\'o}}}{2021}]{konyves-toth:21}
{K{\"o}nyves-T{\'o}th} R.,  {Vink{\'o}} J.,  2021, \mn@doi [\apj]
  {10.3847/1538-4357/abd6c8}, \href
  {https://ui.adsabs.harvard.edu/abs/2021ApJ...909...24K} {909, 24}

\bibitem[\protect\citeauthoryear{{Kumar}, {Narayan}  \& {Johnson}}{{Kumar}
  et~al.}{2008}]{kumar:08}
{Kumar} P.,  {Narayan} R.,   {Johnson} J.~L.,  2008, \mn@doi [\mnras]
  {10.1111/j.1365-2966.2008.13493.x}, \href
  {https://ui.adsabs.harvard.edu/abs/2008MNRAS.388.1729K} {388, 1729}

\bibitem[\protect\citeauthoryear{{Kushnir}, {Zaldarriaga}, {Kollmeier}  \&
  {Waldman}}{{Kushnir} et~al.}{2017}]{kushnir:17}
{Kushnir} D.,  {Zaldarriaga} M.,  {Kollmeier} J.~A.,   {Waldman} R.,  2017,
  \mn@doi [\mnras] {10.1093/mnras/stx255}, \href
  {http://adsabs.harvard.edu/abs/2017MNRAS.467.2146K} {467, 2146}

\bibitem[\protect\citeauthoryear{{Laplace}, {G{\"o}tberg}, {de Mink}, {Justham}
   \& {Farmer}}{{Laplace} et~al.}{2020}]{laplace:20}
{Laplace} E.,  {G{\"o}tberg} Y.,  {de Mink} S.~E.,  {Justham} S.,   {Farmer}
  R.,  2020, \mn@doi [\aap] {10.1051/0004-6361/201937300}, \href
  {https://ui.adsabs.harvard.edu/abs/2020A&A...637A...6L} {637, A6}

\bibitem[\protect\citeauthoryear{{Lin}, {Wang}, {Wang}  \& {Dai}}{{Lin}
  et~al.}{2020}]{lin:20}
{Lin} W.~L.,  {Wang} X.~F.,  {Wang} L.~J.,   {Dai} Z.~G.,  2020, \mn@doi
  [\apjl] {10.3847/2041-8213/abc254}, \href
  {https://ui.adsabs.harvard.edu/abs/2020ApJ...903L..24L} {903, L24}

\bibitem[\protect\citeauthoryear{{Lin}, {Wang}, {Wang}  \& {Dai}}{{Lin}
  et~al.}{2021}]{lin:21}
{Lin} W.,  {Wang} X.,  {Wang} L.,   {Dai} Z.,  2021, \mn@doi [\apjl]
  {10.3847/2041-8213/ac004a}, \href
  {https://ui.adsabs.harvard.edu/abs/2021ApJ...914L...2L} {914, L2}

\bibitem[\protect\citeauthoryear{{Ma} \& {Fuller}}{{Ma} \&
  {Fuller}}{2019}]{ma:19}
{Ma} L.,  {Fuller} J.,  2019, \mn@doi [\mnras] {10.1093/mnras/stz2009}, \href
  {https://ui.adsabs.harvard.edu/abs/2019MNRAS.488.4338M} {488, 4338}

\bibitem[\protect\citeauthoryear{{MacFadyen} \& {Woosley}}{{MacFadyen} \&
  {Woosley}}{1999}]{macfadyen:99}
{MacFadyen} A.~I.,  {Woosley} S.~E.,  1999, \mn@doi [\apj] {10.1086/307790},
  \href {http://adsabs.harvard.edu/abs/1999ApJ...524..262M} {524, 262}

\bibitem[\protect\citeauthoryear{{Maeda}, {Nakamura}, {Nomoto}, {Mazzali},
  {Patat}  \& {Hachisu}}{{Maeda} et~al.}{2002}]{maeda:02}
{Maeda} K.,  {Nakamura} T.,  {Nomoto} K.,  {Mazzali} P.~A.,  {Patat} F.,
  {Hachisu} I.,  2002, \mn@doi [\apj] {10.1086/324487}, \href
  {https://ui.adsabs.harvard.edu/abs/2002ApJ...565..405M} {565, 405}

\bibitem[\protect\citeauthoryear{{Maeda} et~al.,}{{Maeda}
  et~al.}{2007}]{maeda:07}
{Maeda} K.,  et~al., 2007, \mn@doi [\apj] {10.1086/520054}, \href
  {https://ui.adsabs.harvard.edu/abs/2007ApJ...666.1069M} {666, 1069}

\bibitem[\protect\citeauthoryear{{Maeder}}{{Maeder}}{1987}]{maeder:87}
{Maeder} A.,  1987, \aap, \href
  {http://adsabs.harvard.edu/abs/1987A%26A...178..159M} {178, 159}

\bibitem[\protect\citeauthoryear{{Mandel} \& {de Mink}}{{Mandel} \& {de
  Mink}}{2016}]{mandel:16}
{Mandel} I.,  {de Mink} S.~E.,  2016, \mn@doi [\mnras] {10.1093/mnras/stw379},
  \href {http://adsabs.harvard.edu/abs/2016MNRAS.458.2634M} {458, 2634}

\bibitem[\protect\citeauthoryear{{Marchant}, {Langer}, {Podsiadlowski},
  {Tauris}, {de Mink}, {Mandel}  \& {Moriya}}{{Marchant}
  et~al.}{2017}]{marchant:17}
{Marchant} P.,  {Langer} N.,  {Podsiadlowski} P.,  {Tauris} T.~M.,  {de Mink}
  S.,  {Mandel} I.,   {Moriya} T.~J.,  2017, \mn@doi [\aap]
  {10.1051/0004-6361/201630188}, \href
  {http://adsabs.harvard.edu/abs/2017A%26A...604A..55M} {604, A55}

\bibitem[\protect\citeauthoryear{{Marchant}, {Pappas}, {Gallegos-Garcia},
  {Berry}, {Taam}, {Kalogera}  \& {Podsiadlowski}}{{Marchant}
  et~al.}{2021}]{marchant:21}
{Marchant} P.,  {Pappas} K. M.~W.,  {Gallegos-Garcia} M.,  {Berry} C. P.~L.,
  {Taam} R.~E.,  {Kalogera} V.,   {Podsiadlowski} P.,  2021, \mn@doi [\aap]
  {10.1051/0004-6361/202039992}, \href
  {https://ui.adsabs.harvard.edu/abs/2021A&A...650A.107M} {650, A107}

\bibitem[\protect\citeauthoryear{{Margalit}, {Metzger}, {Thompson}, {Nicholl}
  \& {Sukhbold}}{{Margalit} et~al.}{2018}]{margalit:18}
{Margalit} B.,  {Metzger} B.~D.,  {Thompson} T.~A.,  {Nicholl} M.,   {Sukhbold}
  T.,  2018, \mn@doi [\mnras] {10.1093/mnras/sty013}, \href
  {https://ui.adsabs.harvard.edu/abs/2018MNRAS.475.2659M} {475, 2659}

\bibitem[\protect\citeauthoryear{{Mazzali}, {McFadyen}, {Woosley}, {Pian}  \&
  {Tanaka}}{{Mazzali} et~al.}{2014}]{mazzali:14}
{Mazzali} P.~A.,  {McFadyen} A.~I.,  {Woosley} S.~E.,  {Pian} E.,   {Tanaka}
  M.,  2014, \mn@doi [\mnras] {10.1093/mnras/stu1124}, \href
  {https://ui.adsabs.harvard.edu/abs/2014MNRAS.443...67M} {443, 67}

\bibitem[\protect\citeauthoryear{Metzger, Giannios, Thompson, Bucciantini  \&
  Quataert}{Metzger et~al.}{2011}]{metzger:11}
Metzger B.~D.,  Giannios D.,  Thompson T.~A.,  Bucciantini N.,   Quataert E.,
  2011, \mn@doi [Monthly Notices of the Royal Astronomical Society]
  {10.1111/j.1365-2966.2011.18280.x}, 413, 2031

\bibitem[\protect\citeauthoryear{{Metzger}, {Margalit}, {Kasen}  \&
  {Quataert}}{{Metzger} et~al.}{2015}]{metzger:15}
{Metzger} B.~D.,  {Margalit} B.,  {Kasen} D.,   {Quataert} E.,  2015, \mn@doi
  [\mnras] {10.1093/mnras/stv2224}, \href
  {http://adsabs.harvard.edu/abs/2015MNRAS.454.3311M} {454, 3311}

\bibitem[\protect\citeauthoryear{{Metzger}, {Beniamini}  \&
  {Giannios}}{{Metzger} et~al.}{2018}]{metzger:18}
{Metzger} B.~D.,  {Beniamini} P.,   {Giannios} D.,  2018, \mn@doi [\apj]
  {10.3847/1538-4357/aab70c}, \href
  {https://ui.adsabs.harvard.edu/abs/2018ApJ...857...95M} {857, 95}

\bibitem[\protect\citeauthoryear{{Miller} \& {Miller}}{{Miller} \&
  {Miller}}{2015}]{miller:15}
{Miller} M.~C.,  {Miller} J.~M.,  2015, \mn@doi [\physrep]
  {10.1016/j.physrep.2014.09.003}, \href
  {http://adsabs.harvard.edu/abs/2015PhR...548....1M} {548, 1}

\bibitem[\protect\citeauthoryear{{Miller}, {Callister}  \& {Farr}}{{Miller}
  et~al.}{2020}]{miller:20}
{Miller} S.,  {Callister} T.~A.,   {Farr} W.~M.,  2020, \mn@doi [\apj]
  {10.3847/1538-4357/ab80c0}, \href
  {https://ui.adsabs.harvard.edu/abs/2020ApJ...895..128M} {895, 128}

\bibitem[\protect\citeauthoryear{{Modjaz} et~al.,}{{Modjaz}
  et~al.}{2020}]{modjaz:20}
{Modjaz} M.,  et~al., 2020, \mn@doi [\apj] {10.3847/1538-4357/ab4185}, \href
  {https://ui.adsabs.harvard.edu/abs/2020ApJ...892..153M} {892, 153}

\bibitem[\protect\citeauthoryear{{Mosser} et~al.,}{{Mosser}
  et~al.}{2012}]{mosser:12}
{Mosser} B.,  et~al., 2012, \mn@doi [\aap] {10.1051/0004-6361/201220106}, \href
  {http://adsabs.harvard.edu/abs/2012A%26A...548A..10M} {548, A10}

\bibitem[\protect\citeauthoryear{{M{\"u}ller} et~al.,}{{M{\"u}ller}
  et~al.}{2018}]{muller:18}
{M{\"u}ller} B.,  et~al., 2018, preprint, \href
  {http://adsabs.harvard.edu/abs/2018arXiv181105483M} {} (\mn@eprint {arXiv}
  {1811.05483})

\bibitem[\protect\citeauthoryear{{Narayan} \& {Yi}}{{Narayan} \&
  {Yi}}{1994}]{narayan:94}
{Narayan} R.,  {Yi} I.,  1994, \mn@doi [\apjl] {10.1086/187381}, \href
  {https://ui.adsabs.harvard.edu/abs/1994ApJ...428L..13N} {428, L13}

\bibitem[\protect\citeauthoryear{{Narayan}, {Piran}  \& {Kumar}}{{Narayan}
  et~al.}{2001}]{narayan:01}
{Narayan} R.,  {Piran} T.,   {Kumar} P.,  2001, \mn@doi [\apj]
  {10.1086/322267}, \href
  {https://ui.adsabs.harvard.edu/abs/2001ApJ...557..949N} {557, 949}

\bibitem[\protect\citeauthoryear{{Nugis} \& {Lamers}}{{Nugis} \&
  {Lamers}}{2000}]{nugis:00}
{Nugis} T.,  {Lamers} H.~J.~G.~L.~M.,  2000, \aap, \href
  {https://ui.adsabs.harvard.edu/abs/2000A&A...360..227N} {360, 227}

\bibitem[\protect\citeauthoryear{{O'Connor} \& {Ott}}{{O'Connor} \&
  {Ott}}{2011}]{oconnor:11}
{O'Connor} E.,  {Ott} C.~D.,  2011, \mn@doi [\apj]
  {10.1088/0004-637X/730/2/70}, \href
  {https://ui.adsabs.harvard.edu/abs/2011ApJ...730...70O} {730, 70}

\bibitem[\protect\citeauthoryear{{Olejak} \& {Belczynski}}{{Olejak} \&
  {Belczynski}}{2021}]{olejak:21}
{Olejak} A.,  {Belczynski} K.,  2021, arXiv e-prints, \href
  {https://ui.adsabs.harvard.edu/abs/2021arXiv210906872O} {p. arXiv:2109.06872}

\bibitem[\protect\citeauthoryear{{Ouazzani}, {Marques}, {Goupil}, {Christophe},
  {Antoci}  \& {Salmon}}{{Ouazzani} et~al.}{2018}]{ouazzani:18}
{Ouazzani} R.-M.,  {Marques} J.~P.,  {Goupil} M.,  {Christophe} S.,  {Antoci}
  V.,   {Salmon} S.~J.~A.~J.,  2018, preprint, \href
  {http://adsabs.harvard.edu/abs/2018arXiv180109228O} {} (\mn@eprint {arXiv}
  {1801.09228})

\bibitem[\protect\citeauthoryear{{Pandey} et~al.,}{{Pandey}
  et~al.}{2021}]{pandey:21}
{Pandey} S.~B.,  et~al., 2021, arXiv e-prints, \href
  {https://ui.adsabs.harvard.edu/abs/2021arXiv210615856P} {p. arXiv:2106.15856}

\bibitem[\protect\citeauthoryear{{Pavlovskii}, {Ivanova}, {Belczynski}  \&
  {Van}}{{Pavlovskii} et~al.}{2017}]{pavlovskii:17}
{Pavlovskii} K.,  {Ivanova} N.,  {Belczynski} K.,   {Van} K.~X.,  2017, \mn@doi
  [\mnras] {10.1093/mnras/stw2786}, \href
  {https://ui.adsabs.harvard.edu/abs/2017MNRAS.465.2092P} {465, 2092}

\bibitem[\protect\citeauthoryear{{Paxton}, {Bildsten}, {Dotter}, {Herwig},
  {Lesaffre}  \& {Timmes}}{{Paxton} et~al.}{2011}]{paxton:11}
{Paxton} B.,  {Bildsten} L.,  {Dotter} A.,  {Herwig} F.,  {Lesaffre} P.,
  {Timmes} F.,  2011, \mn@doi [\apjs] {10.1088/0067-0049/192/1/3}, \href
  {http://adsabs.harvard.edu/abs/2011ApJS..192....3P} {192, 3}

\bibitem[\protect\citeauthoryear{{Paxton} et~al.,}{{Paxton}
  et~al.}{2013}]{paxton:13}
{Paxton} B.,  et~al., 2013, \mn@doi [\apjs] {10.1088/0067-0049/208/1/4}, \href
  {http://adsabs.harvard.edu/abs/2013ApJS..208....4P} {208, 4}

\bibitem[\protect\citeauthoryear{{Paxton} et~al.,}{{Paxton}
  et~al.}{2015}]{paxton:15}
{Paxton} B.,  et~al., 2015, \mn@doi [\apjs] {10.1088/0067-0049/220/1/15}, \href
  {http://adsabs.harvard.edu/abs/2015ApJS..220...15P} {220, 15}

\bibitem[\protect\citeauthoryear{{Paxton} et~al.,}{{Paxton}
  et~al.}{2018}]{paxton:18}
{Paxton} B.,  et~al., 2018, \mn@doi [\apjs] {10.3847/1538-4365/aaa5a8}, \href
  {http://adsabs.harvard.edu/abs/2018ApJS..234...34P} {234, 34}

\bibitem[\protect\citeauthoryear{{Paxton} et~al.,}{{Paxton}
  et~al.}{2019}]{paxton:19}
{Paxton} B.,  et~al., 2019, arXiv e-prints, \href
  {http://adsabs.harvard.edu/abs/2019arXiv190301426P} {}

\bibitem[\protect\citeauthoryear{{Perley} et~al.,}{{Perley}
  et~al.}{2016}]{perley:16}
{Perley} D.~A.,  et~al., 2016, \mn@doi [\apj] {10.3847/0004-637X/817/1/8},
  \href {http://adsabs.harvard.edu/abs/2016ApJ...817....8P} {817, 8}

\bibitem[\protect\citeauthoryear{{Piro} \& {Ott}}{{Piro} \&
  {Ott}}{2011}]{piro:11}
{Piro} A.~L.,  {Ott} C.~D.,  2011, \mn@doi [\apj]
  {10.1088/0004-637X/736/2/108}, \href
  {https://ui.adsabs.harvard.edu/abs/2011ApJ...736..108P} {736, 108}

\bibitem[\protect\citeauthoryear{{Popham}, {Woosley}  \& {Fryer}}{{Popham}
  et~al.}{1999}]{popham:99}
{Popham} R.,  {Woosley} S.~E.,   {Fryer} C.,  1999, \mn@doi [\apj]
  {10.1086/307259}, \href
  {https://ui.adsabs.harvard.edu/abs/1999ApJ...518..356P} {518, 356}

\bibitem[\protect\citeauthoryear{{Popov} \& {Turolla}}{{Popov} \&
  {Turolla}}{2012}]{popov:12}
{Popov} S.~B.,  {Turolla} R.,  2012, \mn@doi [\apss]
  {10.1007/s10509-012-1100-z}, \href
  {http://adsabs.harvard.edu/abs/2012Ap%26SS.341..457P} {341, 457}

\bibitem[\protect\citeauthoryear{{Popov}, {Pons}, {Miralles}, {Boldin}  \&
  {Posselt}}{{Popov} et~al.}{2010}]{popov:10}
{Popov} S.~B.,  {Pons} J.~A.,  {Miralles} J.~A.,  {Boldin} P.~A.,   {Posselt}
  B.,  2010, \mn@doi [\mnras] {10.1111/j.1365-2966.2009.15850.x}, \href
  {http://adsabs.harvard.edu/abs/2010MNRAS.401.2675P} {401, 2675}

\bibitem[\protect\citeauthoryear{{Qin}, {Fragos}, {Meynet}, {Andrews},
  {S{\o}rensen}  \& {Song}}{{Qin} et~al.}{2018}]{qin:18}
{Qin} Y.,  {Fragos} T.,  {Meynet} G.,  {Andrews} J.,  {S{\o}rensen} M.,
  {Song} H.~F.,  2018, \mn@doi [\aap] {10.1051/0004-6361/201832839}, \href
  {http://adsabs.harvard.edu/abs/2018A%26A...616A..28Q} {616, A28}

\bibitem[\protect\citeauthoryear{{Qin}, {Marchant}, {Fragos}, {Meynet}  \&
  {Kalogera}}{{Qin} et~al.}{2019}]{qin:19}
{Qin} Y.,  {Marchant} P.,  {Fragos} T.,  {Meynet} G.,   {Kalogera} V.,  2019,
  \mn@doi [\apjl] {10.3847/2041-8213/aaf97b}, \href
  {https://ui.adsabs.harvard.edu/abs/2019ApJ...870L..18Q} {870, L18}

\bibitem[\protect\citeauthoryear{{Roulet}, {Chia}, {Olsen}, {Dai},
  {Venumadhav}, {Zackay}  \& {Zaldarriaga}}{{Roulet} et~al.}{2021}]{roulet:21}
{Roulet} J.,  {Chia} H.~S.,  {Olsen} S.,  {Dai} L.,  {Venumadhav} T.,  {Zackay}
  B.,   {Zaldarriaga} M.,  2021, arXiv e-prints, \href
  {https://ui.adsabs.harvard.edu/abs/2021arXiv210510580R} {p. arXiv:2105.10580}

\bibitem[\protect\citeauthoryear{{Schr{\o}der}, {Batta}  \&
  {Ramirez-Ruiz}}{{Schr{\o}der} et~al.}{2018}]{schroder:18}
{Schr{\o}der} S.~L.,  {Batta} A.,   {Ramirez-Ruiz} E.,  2018, \mn@doi [\apjl]
  {10.3847/2041-8213/aacf8d}, \href
  {https://ui.adsabs.harvard.edu/abs/2018ApJ...862L...3S} {862, L3}

\bibitem[\protect\citeauthoryear{{Shakura} \& {Sunyaev}}{{Shakura} \&
  {Sunyaev}}{1973}]{shakura:73}
{Shakura} N.~I.,  {Sunyaev} R.~A.,  1973, \aap, \href
  {https://ui.adsabs.harvard.edu/abs/1973A&A....24..337S} {500, 33}

\bibitem[\protect\citeauthoryear{{Shankar}, {M{\"o}sta}, {Barnes}, {Duffell}
  \& {Kasen}}{{Shankar} et~al.}{2021}]{shankar:21}
{Shankar} S.,  {M{\"o}sta} P.,  {Barnes} J.,  {Duffell} P.~C.,   {Kasen} D.,
  2021, arXiv e-prints, \href
  {https://ui.adsabs.harvard.edu/abs/2021arXiv210508092S} {p. arXiv:2105.08092}

\bibitem[\protect\citeauthoryear{{Shivvers} et~al.,}{{Shivvers}
  et~al.}{2017}]{shivvers:17b}
{Shivvers} I.,  et~al., 2017, \mn@doi [\pasp] {10.1088/1538-3873/aa54a6}, \href
  {https://ui.adsabs.harvard.edu/abs/2017PASP..129e4201S} {129, 054201}

\bibitem[\protect\citeauthoryear{{Siegel}, {Barnes}  \& {Metzger}}{{Siegel}
  et~al.}{2019}]{siegel:19}
{Siegel} D.~M.,  {Barnes} J.,   {Metzger} B.~D.,  2019, \mn@doi [\nat]
  {10.1038/s41586-019-1136-0}, \href
  {https://ui.adsabs.harvard.edu/abs/2019Natur.569..241S} {569, 241}

\bibitem[\protect\citeauthoryear{{Spada}, {Gellert}, {Arlt}  \&
  {Deheuvels}}{{Spada} et~al.}{2016}]{spada:16}
{Spada} F.,  {Gellert} M.,  {Arlt} R.,   {Deheuvels} S.,  2016, \mn@doi [\aap]
  {10.1051/0004-6361/201527591}, \href
  {http://adsabs.harvard.edu/abs/2016A%26A...589A..23S} {589, A23}

\bibitem[\protect\citeauthoryear{{Spruit}}{{Spruit}}{1999}]{spruit:99}
{Spruit} H.~C.,  1999, \aap, \href
  {http://adsabs.harvard.edu/abs/1999A%26A...349..189S} {349, 189}

\bibitem[\protect\citeauthoryear{{Spruit}}{{Spruit}}{2002}]{spruit:02}
{Spruit} H.~C.,  2002, \mn@doi [\aap] {10.1051/0004-6361:20011465}, \href
  {http://adsabs.harvard.edu/abs/2002A%26A...381..923S} {381, 923}

\bibitem[\protect\citeauthoryear{{Stockinger} et~al.,}{{Stockinger}
  et~al.}{2020}]{stockinger:20}
{Stockinger} G.,  et~al., 2020, \mn@doi [\mnras] {10.1093/mnras/staa1691},
  \href {https://ui.adsabs.harvard.edu/abs/2020MNRAS.496.2039S} {496, 2039}

\bibitem[\protect\citeauthoryear{{Takahashi} \& {Langer}}{{Takahashi} \&
  {Langer}}{2021}]{takahashi:21}
{Takahashi} K.,  {Langer} N.,  2021, \mn@doi [\aap]
  {10.1051/0004-6361/202039253}, \href
  {https://ui.adsabs.harvard.edu/abs/2021A&A...646A..19T} {646, A19}

\bibitem[\protect\citeauthoryear{{Tauris}, {Langer}  \&
  {Podsiadlowski}}{{Tauris} et~al.}{2015}]{tauris:15}
{Tauris} T.~M.,  {Langer} N.,   {Podsiadlowski} P.,  2015, \mn@doi [\mnras]
  {10.1093/mnras/stv990}, \href
  {https://ui.adsabs.harvard.edu/abs/2015MNRAS.451.2123T} {451, 2123}

\bibitem[\protect\citeauthoryear{{Tauris} et~al.,}{{Tauris}
  et~al.}{2017}]{tauris:17}
{Tauris} T.~M.,  et~al., 2017, \mn@doi [\apj] {10.3847/1538-4357/aa7e89}, \href
  {https://ui.adsabs.harvard.edu/abs/2017ApJ...846..170T} {846, 170}

\bibitem[\protect\citeauthoryear{{Tayar}, {Beck}, {Pinsonneault}, {Garc{\'\i}a}
   \& {Mathur}}{{Tayar} et~al.}{2019}]{tayar:19}
{Tayar} J.,  {Beck} P.~G.,  {Pinsonneault} M.~H.,  {Garc{\'\i}a} R.~A.,
  {Mathur} S.,  2019, arXiv e-prints, \href
  {https://ui.adsabs.harvard.edu/abs/2019arXiv191101443T} {p. arXiv:1911.01443}

\bibitem[\protect\citeauthoryear{{The LIGO Scientific Collaboration}
  et~al.,}{{The LIGO Scientific Collaboration} et~al.}{2018}]{ligoo2b:18}
{The LIGO Scientific Collaboration} et~al., 2018, arXiv e-prints, \href
  {http://adsabs.harvard.edu/abs/2018arXiv181112940T} {}

\bibitem[\protect\citeauthoryear{{Triana}, {Corsaro}, {De Ridder}, {Bonanno},
  {P{\'e}rez Hern{\'a}ndez}  \& {Garc{\'{\i}}a}}{{Triana}
  et~al.}{2017}]{triana:17}
{Triana} S.~A.,  {Corsaro} E.,  {De Ridder} J.,  {Bonanno} A.,  {P{\'e}rez
  Hern{\'a}ndez} F.,   {Garc{\'{\i}}a} R.~A.,  2017, \mn@doi [\aap]
  {10.1051/0004-6361/201629186}, \href
  {http://adsabs.harvard.edu/abs/2017A%26A...602A..62T} {602, A62}

\bibitem[\protect\citeauthoryear{{Vink}, {de Koter}  \& {Lamers}}{{Vink}
  et~al.}{2000}]{vink:00}
{Vink} J.~S.,  {de Koter} A.,   {Lamers} H.~J.~G.~L.~M.,  2000, \aap, \href
  {https://ui.adsabs.harvard.edu/abs/2000A&A...362..295V} {362, 295}

\bibitem[\protect\citeauthoryear{{Vink}, {de Koter}  \& {Lamers}}{{Vink}
  et~al.}{2001}]{vink:01}
{Vink} J.~S.,  {de Koter} A.,   {Lamers} H.~J.~G.~L.~M.,  2001, \mn@doi [\aap]
  {10.1051/0004-6361:20010127}, \href
  {https://ui.adsabs.harvard.edu/abs/2001A&A...369..574V} {369, 574}

\bibitem[\protect\citeauthoryear{{Wanderman} \& {Piran}}{{Wanderman} \&
  {Piran}}{2010}]{wanderman:10}
{Wanderman} D.,  {Piran} T.,  2010, \mn@doi [\mnras]
  {10.1111/j.1365-2966.2010.16787.x}, \href
  {http://adsabs.harvard.edu/abs/2010MNRAS.406.1944W} {406, 1944}

\bibitem[\protect\citeauthoryear{{Woosley}}{{Woosley}}{1993}]{woosley:93}
{Woosley} S.~E.,  1993, \mn@doi [\apj] {10.1086/172359}, \href
  {http://adsabs.harvard.edu/abs/1993ApJ...405..273W} {405, 273}

\bibitem[\protect\citeauthoryear{Woosley \& Heger}{Woosley \&
  Heger}{2006}]{woosley:06}
Woosley S.~E.,  Heger A.,  2006, \mn@doi [{ApJ}] {10.1086/498500}, 637, 914

\bibitem[\protect\citeauthoryear{{Woosley}, {Heger}  \& {Weaver}}{{Woosley}
  et~al.}{2002}]{woosley:02}
{Woosley} S.~E.,  {Heger} A.,   {Weaver} T.~A.,  2002, \mn@doi [Reviews of
  Modern Physics] {10.1103/RevModPhys.74.1015}, \href
  {http://adsabs.harvard.edu/abs/2002RvMP...74.1015W} {74, 1015}

\bibitem[\protect\citeauthoryear{{Worley}, {Krastev}  \& {Li}}{{Worley}
  et~al.}{2008}]{worley:08}
{Worley} A.,  {Krastev} P.~G.,   {Li} B.-A.,  2008, \mn@doi [\apj]
  {10.1086/589823}, \href
  {https://ui.adsabs.harvard.edu/abs/2008ApJ...685..390W} {685, 390}

\bibitem[\protect\citeauthoryear{Yoon, Langer  \& Norman}{Yoon
  et~al.}{2006}]{yoon:06}
Yoon S.-C.,  Langer N.,   Norman C.,  2006, \mn@doi [Astronomy and
  Astrophysics] {10.1051/0004-6361:20065912}, 460, 199

\bibitem[\protect\citeauthoryear{{Yoon}, {Woosley}  \& {Langer}}{{Yoon}
  et~al.}{2010}]{yoon:10}
{Yoon} S.~C.,  {Woosley} S.~E.,   {Langer} N.,  2010, \mn@doi [\apj]
  {10.1088/0004-637X/725/1/940}, \href
  {https://ui.adsabs.harvard.edu/abs/2010ApJ...725..940Y} {725, 940}

\bibitem[\protect\citeauthoryear{{Yu}, {Zhu}, {Li}, {L{\"u}}  \& {Zou}}{{Yu}
  et~al.}{2017}]{yu:17}
{Yu} Y.-W.,  {Zhu} J.-P.,  {Li} S.-Z.,  {L{\"u}} H.-J.,   {Zou} Y.-C.,  2017,
  \mn@doi [\apj] {10.3847/1538-4357/aa6c27}, \href
  {https://ui.adsabs.harvard.edu/abs/2017ApJ...840...12Y} {840, 12}

\bibitem[\protect\citeauthoryear{{Yuan} \& {Narayan}}{{Yuan} \&
  {Narayan}}{2014}]{yuan:14}
{Yuan} F.,  {Narayan} R.,  2014, \mn@doi [\araa]
  {10.1146/annurev-astro-082812-141003}, \href
  {https://ui.adsabs.harvard.edu/abs/2014ARA&A..52..529Y} {52, 529}

\bibitem[\protect\citeauthoryear{{Zahn}}{{Zahn}}{1977}]{zahn:77}
{Zahn} J.~P.,  1977, \aap, \href
  {https://ui.adsabs.harvard.edu/abs/1977A&A....57..383Z} {500, 121}

\bibitem[\protect\citeauthoryear{{Zaldarriaga}, {Kushnir}  \&
  {Kollmeier}}{{Zaldarriaga} et~al.}{2018}]{zaldarriaga:18}
{Zaldarriaga} M.,  {Kushnir} D.,   {Kollmeier} J.~A.,  2018, \mn@doi [\mnras]
  {10.1093/mnras/stx2577}, \href
  {http://adsabs.harvard.edu/abs/2018MNRAS.473.4174Z} {473, 4174}

\bibitem[\protect\citeauthoryear{{Zapartas} et~al.,}{{Zapartas}
  et~al.}{2021}]{zapartas:21}
{Zapartas} E.,  et~al., 2021, arXiv e-prints, \href
  {https://ui.adsabs.harvard.edu/abs/2021arXiv210605228Z} {p. arXiv:2106.05228}

\bibitem[\protect\citeauthoryear{{Zevin} et~al.,}{{Zevin}
  et~al.}{2021}]{zevin:21}
{Zevin} M.,  et~al., 2021, \mn@doi [\apj] {10.3847/1538-4357/abe40e}, \href
  {https://ui.adsabs.harvard.edu/abs/2021ApJ...910..152Z} {910, 152}

\bibitem[\protect\citeauthoryear{{Zhao}, {Xue}  \& {Cao}}{{Zhao}
  et~al.}{2021}]{zhao:21}
{Zhao} W.-C.,  {Xue} X.-X.,   {Cao} X.-F.,  2021, \mn@doi [New Astronomy]
  {10.1016/j.newast.2020.101506}, \href
  {https://ui.adsabs.harvard.edu/abs/2021NewA...8301506Z} {83, 101506}

\bibitem[\protect\citeauthoryear{{de Jager}, {Nieuwenhuijzen}  \& {van der
  Hucht}}{{de Jager} et~al.}{1988}]{dejager:88}
{de Jager} C.,  {Nieuwenhuijzen} H.,   {van der Hucht} K.~A.,  1988, \aaps,
  \href {https://ui.adsabs.harvard.edu/abs/1988A&AS...72..259D} {72, 259}

\bibitem[\protect\citeauthoryear{{de Mink} \& {Mandel}}{{de Mink} \&
  {Mandel}}{2016}]{demink:16}
{de Mink} S.~E.,  {Mandel} I.,  2016, \mn@doi [\mnras] {10.1093/mnras/stw1219},
  \href {http://adsabs.harvard.edu/abs/2016MNRAS.460.3545D} {460, 3545}

\bibitem[\protect\citeauthoryear{{van Son} et~al.,}{{van Son}
  et~al.}{2021}]{vanson:21}
{van Son} L.~A.~C.,  et~al., 2021, arXiv e-prints, \href
  {https://ui.adsabs.harvard.edu/abs/2021arXiv211001634V} {p. arXiv:2110.01634}

\makeatother
\end{thebibliography}

\appendix

\section{Accretion Energetics}
\label{accretion}

A crude one-zone model for the BH accretion disk and its time evolution is provided by \cite{kumar:08}, who included the combined actions of mass infall, viscous accretion, and disk wind. We adopt their model and briefly describe the procedure here\footnote{Our code that models the BH accretion history can be downloaded from this URL: https://github.com/wenbinlu/collapsar.git}. At a given moment, the disk has mass $M_{\rm d}$ and AM $J_{\rm d}$, so the characteristic radius is given by $r_{\rm d}=J_{\rm d}^2/(GM_{\rm BH}M_{\rm d}^2)$ assuming Keplerian rotation. The time evolution of these two quantities are given by
  \beq
  \dot{M}_{\rm d} = \dot{M}_{\rm fb} - {M_{\rm d}\over t_{\rm vis}},\
  \eeq
  \beq
  \dot{J}_{\rm d} = \dot{J}_{\rm fb} - \dot{M}_{\rm BH} j_{\rm ISCO} - f_{\rm AM} {J_{\rm d}\over t_{\rm vis}},
  \eeq
where $\dot{M}_{\rm fb}$ and $\dot{J}_{\rm fb}$ are the mass and AM fallback rates of the envelope, $t_{\rm vis}=\alpha_{\rm s}^{-1} \sqrt{r_{\rm d}^3/GM_{\rm BH}}$ is the viscous timescale, $\alpha_{\rm s}\in(0.01, 0.1)$ is the dimensionless viscous parameter \citep{shakura:73}, and $\dot{M}_{\rm BH}$ is the BH accretion rate at the inner edge of the disk. The parameter $f_{\rm AM}$ relates the specific AM of the mass lost through the disk wind to that of the disk, as described below.

To compute the mass/AM fallback rates from our progenitor model, we first calculate the free-fall time for each shell at radius $r$ as $t_{\rm ff}(r) = (\pi/2^{1/5}) \sqrt{r^3/GM(r)}$, where $M(r)$ is the enclosed mass. Then, the mass fallback rate is given by
\begin{equation}
    \dot{M}_{\rm fb} = \frac{dM}{dt_{\rm ff}} \, ,
\end{equation}
\begin{equation}
    \dot{J}_{\rm fb} = \frac{2}{3} \Omega(r) r^2 \dot{M}_{\rm fb}.
\end{equation}
These are the rates at which mass and AM are added to either the BH or the accretion disk at a given time $t = t_{\rm ff}(r)$. For the disk-forming radial shells with $j>j_{\rm ISCO}$, we assume that only the regions $>30^{\rm o}$ from the rotational axis manage to fallback to the disk whereas the regions with polar angles $<30^{\rm o}$ is removed from the star by the mechanical feedback of the accretion disk wind/jet. This means that the mass and AM fallback rate are slightly reduced by a factor of $\cos 30^{o} = 0.866$ and $(3/2)[\cos 30^{o} - (1/3) (\cos 30^{o})^3] = 0.974$, respectively.

The key ingredient of the disk model is that at a given moment, the disk has a power-law radial mass accretion rate profile $\dot{M}_{\rm acc}(r) = (M_{\rm d}/t_{\rm vis})(r/r_{\rm d})^s$ \citep{blandford:99}, where $s\in(0, 1)$ is a free parameter (likely between 0.3 and 0.8, \citealt{yuan:14}). The power-law flattens below a transition radius $r_{\rm t}$, where the disk starts to be neutrino-cooled, and there is no disk wind from radii $r<r_{\rm t}$, i.e., $\dot{M}_{\rm acc}(r<r_{\rm t})=\dot{M}_{\rm acc}(r_{\rm t})$. The transition radius is given by $\dot{M}_{\rm acc}(r_{\rm t})=\dot{M}_{\rm ign} r_{\rm t}/r_{\rm s}$ ($r_{\rm s}=2GM_{\rm BH}/c^2$ being the Schwarzschild radius), where $\dot{M}_{\rm ign}$ is the critical accretion for URCA ignition that depends on the viscous parameter $\alpha_{\rm s}$, BH mass and spin (see \citealt{chen:07,siegel:19}). We adopt $\alpha_{\rm s}=0.03$ and keep $\dot{M}_{\rm ign}\simeq 10^{-2.5}\,M_\odot\rm\,yr^{-1}$ fixed in this paper. If $r_{\rm t}$ is in between $r_{\rm ISCO}$ and $r_{\rm d}$, then the transition from ADAF to the neutrino-cooled regime indeed occurs, otherwise the radial profile of the accretion rate is either a single power-law (if $r_{\rm t}<r_{\rm ISCO}$) or constant function ($r_{\rm t}>r_{\rm d}$). We further assume that the specific AM of the disk wind is equal to the local Keplerian value of $\sqrt{GM_{\rm BH}r}$ at each radius $r$, which gives $f_{\rm AM}=(1+0.5/s)^{-1}[1 - (r_{\rm t}/r_{\rm d})^{s+0.5}]$.  Then, the mass and AM accretion rates of the BH are $\dot{M}_{\rm BH}=\dot{M}_{\rm acc}(r_{\rm isco})$ and $\dot{J}_{\rm BH}=j_{\rm ISCO}\dot{M}_{\rm BH}$. The total accretion luminosity of the disk is given by
  \beq
  L_{\rm ac}\simeq \eta_{\rm NT} \dot{M}_{\rm BH} c^2,
  \eeq
where the Novikov-Thorne efficiency $\eta_{\rm NT}=1-\sqrt{1-r_{\rm s}/3r_{\rm isco}}$ is given by the energy difference between a particle at infinity and at the ISCO. Under the assumption that each wind fluid element lifted from radius $r$ carries positive specific energy of $GM_{\rm BH}/2r$, we estimate the total mechanical power of the disk wind to be
  \beq
  L_{\rm wind}\simeq {s\over 2(1-s)} {GM_{\rm BH}\over r_{\rm d}} {M_{\rm d}\over t_{\rm vis}}\left[(r_{\rm d}/r_{\rm t})^{1-s} - 1\right].
  \eeq
Finally, we integrate over the entire evolution history and obtain the total energy generated by the accretion disk $E_{\rm ac} = \int L_{\rm ac} dt$ as well as the mechanical energy carried by the disk wind $E_{\rm wind} = \int L_{\rm wind} dt$.  Our nominal result in \autoref{Eengine} uses $s=0.5$, and for other choices of $s\in (0.3, 0.8)$, $E_{\rm wind}$ differs by a factor of a few.

\section{Model Testing}
\label{testing}

\subsection{Resolution Testing}

\begin{figure}
\includegraphics[scale=0.33]{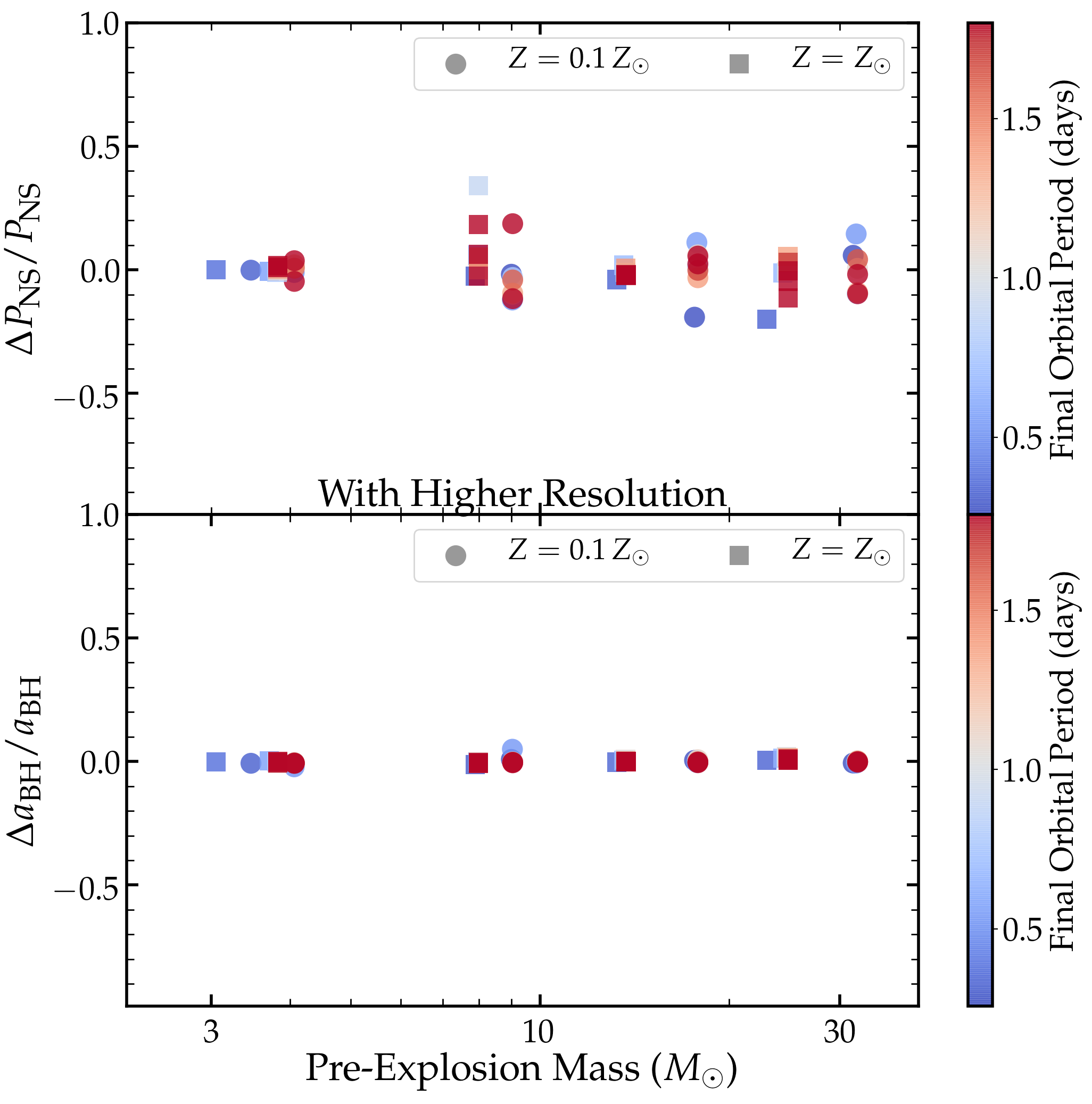}
\caption{\label{fig:testingfig} Fractional difference in the predicted NS rotation period (top) and BH spin (bottom) between models with our highest spatial resolution and slightly lower resolution.
}
\end{figure}

\begin{figure}
\includegraphics[scale=0.33]{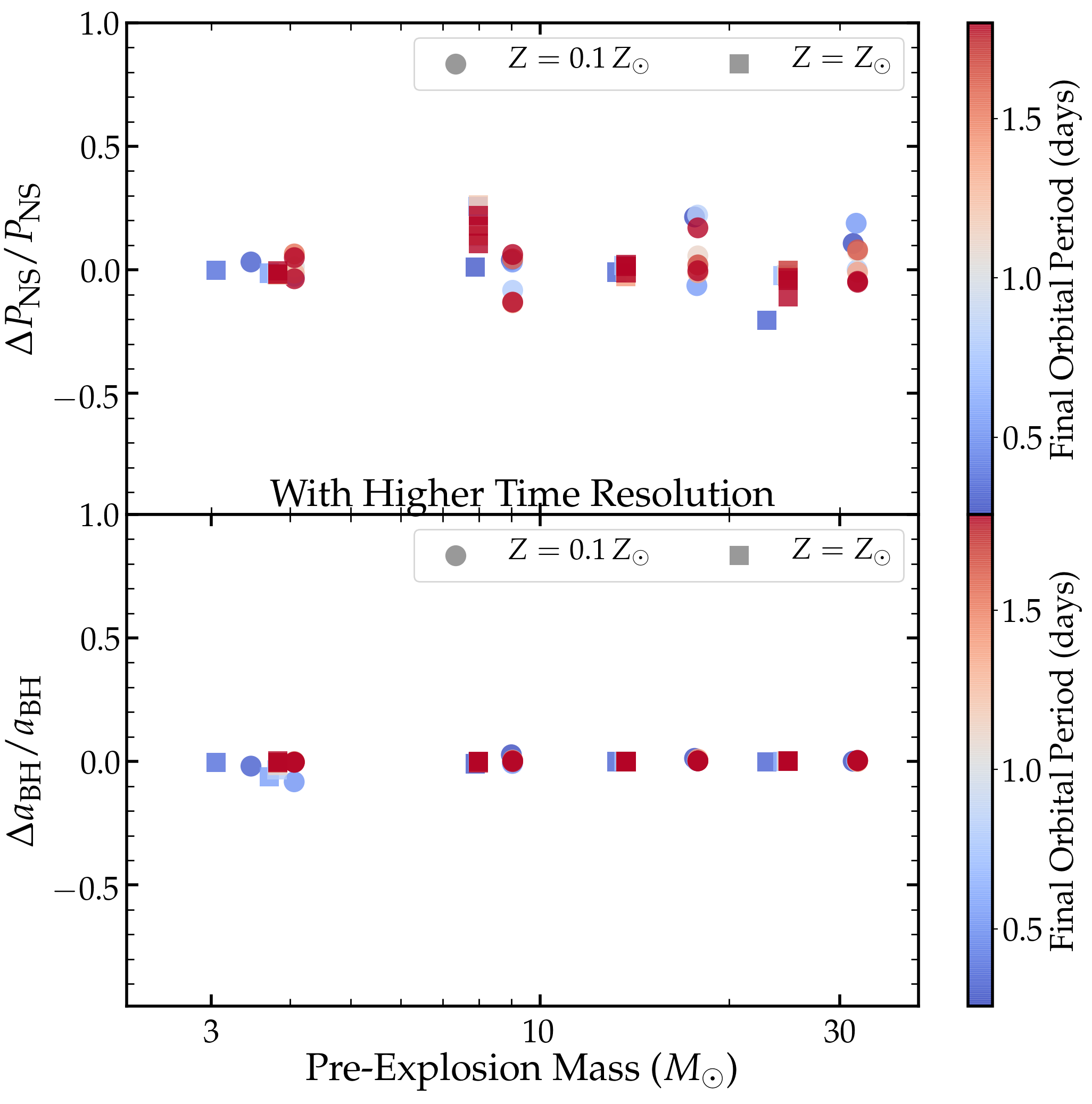}
\caption{\label{fig:testingfig2} Same as \autoref{fig:testingfig} but for models with different time resolution.
}
\end{figure}

It is important to check that our results are not spurious due to unstable or non-converged stellar models. To do this, we performed a few tests. First, we increased each model's spatial resolution by a factor of $\sim$2 by decreasing \verb|mesh_delta_coeff| and \verb|mesh_delta_coeff_for_highT|. \autoref{fig:testingfig} shows that the predicted NS and BH rotation rates for all the models were very similar, within 10\% in most cases and never varying by more than 40\%. The differences appeared to be stochastic, with no systematic shift to higher or lower spins.

Second, we increased each model's time resolution by requiring smaller changes in grid cell temperatures and chemical abundances by a factor of two. The corresponding differences in NS and BH spins are shown in \autoref{fig:testingfig2}. Again, no large or systematic changes were observed. Since the uncertainties in physical prescriptions (see below) are much larger than differences arising from numerical resolution, we consider the models to be sufficiently converged for our purposes.

\end{document}